\documentclass[journal,onecolumn]{IEEEtran}
\usepackage{lineno}
\usepackage{graphicx,float}
\usepackage{amsmath,amssymb}
\usepackage{url}
\usepackage{ifthen}
\usepackage[usenames]{color}
\usepackage{verbatim}
\usepackage{comment}
\usepackage{enumitem}
\usepackage{multirow}
\usepackage[bookmarks,bookmarksdepth=2]{hyperref}
\usepackage{algorithm,algpseudocode}
\newtheorem{proposition}{Proposition}
\newtheorem{theorem}[proposition]{Theorem}

\newtheorem{lemma}[proposition]{Lemma}

\newtheorem{definition}{Definition}
\newtheorem{example}{Example}

\newtheorem{remark}{Remark}
\algrenewcommand\algorithmicrequire{\textbf{Input:}}
\algrenewcommand\algorithmicensure{\textbf{Output:}}

\floatstyle{boxed}
\restylefloat{figure}

\newcommand{\F}{\mathbb{F}}
\newcommand{\lr}{\stackrel{\$}{\leftarrow}}
\newcommand{\Encode}{{\sf Encode}}

\newcommand{\itob}{{\sf I2B}}
\newcommand{\btoi}{{\sf B2I}}

\newcommand{\Gmc}{{\bf G}}
\newcommand{\Pmc}{{\bf P}}
\newcommand{\Smc}{{\bf S}}
\newcommand{\pkmc}{\mathsf{pk_{ME}}}
\newcommand{\skmc}{\mathsf{sk_{ME}}}

\newcommand{\mcsetup}{\mathsf{ME.Setup}}

\newcommand{\mckeygen}{\mathsf{ME.KeyGen}}
\newcommand{\mcenc}{\mathsf{ME.Enc}}
\newcommand{\mcdec}{\mathsf{ME.Dec}}
\newcommand{\nmc}{n}
\newcommand{\kmc}{k}
\newcommand{\tmc}{t}

\newcommand{\ncfs}{m}
\newcommand{\kcfs}{r}
\newcommand{\tcfs}{\omega}
\newcommand{\y}{\mathbf{y}}
\newcommand{\Hcfs}{{\bf H}}

\newcommand{\hash}{\mathcal{H}}

\newcommand{\decode}{{\it Decode}}

\newcommand{\CMT}{{\sf CMT}}
\newcommand{\Ch}{{\sf Ch}}

\newcommand{\RSP}{{\sf RSP}}

\newcommand{\eg}{\emph{e.g.}}
\newcommand{\ie}{\emph{i.e.}}
\begin{document}
	
\title{Provably Secure Group Signature Schemes from Code-Based Assumptions}
	
\author{Martianus Frederic Ezerman, Hyung Tae Lee, San Ling, Khoa Nguyen, and Huaxiong Wang
\thanks{M.~F.~Ezerman, S.~Ling, K.~Nguyen, and H.~Wang are with the School of Physical and Mathematical Sciences, Nanyang Technological University, 21 Nanyang Link, Singapore 637371, 
	e-mails: ${\rm\{fredezerman,lingsan,khoantt,HXWang\}}$@ntu.edu.sg.}
\thanks{H.~T.~Lee is with Division of Computer Science and Engineering, College of Engineering, Chonbuk National University, Republic of Korea, e-mail: ${\rm hyungtaelee}$@chonbuk.ac.kr. H.~T.~Lee is the corresponding author.}
\thanks{
The research was funded by Research Grant TL-9014101684-01 as well as by the Singapore Ministry of
Education's Research Grants MOE2013-T2-1-041 and MOE2016-T2-2-014(S). Hyung Tae Lee was also supported by the National Research Foundation (NRF) grant funded by the Korea government~(MSIT) (No.~NRF-2018R1C1B6008476). Khoa Nguyen was also supported by the Gopalakrishnan--NTU Presidential Postdoctoral Fellowship 2018. Huaxiong Wang was also supported by the National Research Foundation, Singapore Prime Minister’s Office, under its Strategic Capability Research Centres Funding Initiative.
}
\thanks{The present paper is the full extension of our earlier work~\cite{ELLNW15}, in the proceedings of \emph{ASIACRYPT 2015}, that contains the basic scheme.}
\thanks{This work has been submitted to the IEEE for possible publication. Copyright may be transferred without notice, after which this version may no longer be accessible.}
}
\maketitle

\begin{abstract}
We solve an open question in code-based cryptography by introducing two provably secure group signature schemes from code-based assumptions. Our basic scheme satisfies the \textsf{CPA}-anonymity and traceability requirements in the random oracle model, assuming the hardness of the McEliece problem, the Learning Parity with Noise problem, and a variant of the Syndrome Decoding problem. The construction produces smaller key and signature sizes than the previous group signature schemes from lattices, as long as the cardinality of the underlying group does not exceed $2^{24}$, which is roughly comparable to the current population of the Netherlands.
We develop the basic scheme further to achieve the strongest anonymity notion, \ie, \textsf{CCA}-anonymity, with a small overhead in terms of efficiency. 
The feasibility of two proposed schemes is supported by implementation results. 
Our two schemes are the first in their respective classes of provably secure groups signature schemes. Additionally, the techniques introduced in this work might be of independent interest. These are a new verifiable encryption protocol for the randomized McEliece encryption and a novel approach to design formal security reductions from the Syndrome Decoding problem.
\end{abstract}

\begin{IEEEkeywords}
post-quantum cryptography, code-based group signature, zero-knowledge protocol, McEliece encryption, syndrome decoding.
\end{IEEEkeywords}
\section{Introduction}\label{Sec:Introduction}
\subsection{Background and Motivation}
Group signature~\cite{CH91} is a fundamental cryptographic primitive with two intriguing features. The first one is \emph{anonymity}. It allows users of a group to anonymously sign documents on behalf of the whole group. The second one is \emph{traceability}. There exists a tracing authority that can tie a given signature to the signer's identity should the need arise. These two properties make group signatures highly useful in various real-life scenarios such as controlled anonymous printing services, digital right management systems, e-bidding and e-voting schemes. Theoretically, designing secure and efficient group signature schemes is of deep interest since doing so typically requires a sophisticated combination of carefully chosen cryptographic ingredients. Numerous constructions of group signatures have been proposed. Most of them, \eg, the respective schemes in \cite{CS97,ACJT00,BBS04,BW06}, and \cite{LPY12a}, are based on classical number-theoretic assumptions.

While number-theoretic-based group signatures, such as those in \cite{ACJT00} and \cite{BBS04}, can be very efficient, they would become insecure once the era of scalable quantum computing arrives~\cite{Shor97}. Prior to our work, the search for {group signatures that have the potential to be secure against quantum computers}, as a preparation for the future, has been quite active, with at least six published schemes \cite{GKV10,CNR12,LLLS13-Asiacrypt,LLNW14-PKC,LNW15,NZZ15}. All of them are based on computational assumptions from lattices. Despite their theoretical interest, the schemes require significantly large key and signature sizes. None of them has been supported by implementation results. Our evaluation, in Section~\ref{subsec:Our-contributions} below, shows that these lattice-based schemes are indeed very far from being practical. This somewhat unsatisfactory situation highlights two interesting challenges. The first one is to push {group signatures from quantum-resistant assumptions} closer to practice. The second one is to bring more diversity in, with schemes from other candidates for post-quantum cryptography, \eg, code-based, hash-based, and multivariate-based. An easy-to-implement and competitively efficient code-based group signature scheme, for instance, would be highly desirable.

A code-based group signature, in the strongest security model for static groups as discussed in~\cite{BMW03}, typically requires the following three cryptographic layers.
\begin{enumerate}
	\item The first layer requires a {\bf secure (standard) signature scheme} to sign messages. Note that in most schemes based on the model in~\cite{BMW03}, a standard signature is also employed to issue the users' secret keys. However, this is not necessary. The scheme constructed in this paper is an illustrative example. We observe that existing code-based signatures fall into two categories.
	
	The ``hash-and-sign'' category consists of the CFS signature~\cite{CFS01} and its modified versions~\cite{Dal07,Finiasz10,MVR12}. The known security proofs for schemes in this category, however, should be viewed with skepticism. The assumption used in~\cite{Dal07}, for example, had been invalidated by the distinguishing attacks detailed in~\cite{FGOPT13}, while the new assumption proposed in~\cite{MVR12} lies on a rather fragile ground.
		
	The ``Fiat-Shamir'' category consists of schemes derived from Stern's identification protocol in~\cite{Ste96} and its variants in~\cite{Veron96,CVA10}, and \cite{AGS11} via the Fiat-Shamir transformation from~\cite{FS86}. Although these schemes produce relatively large signatures, as the underlying protocol has to be repeated many times to make the soundness error negligibly small, their provable security, in the random oracle model, is well-understood.
	
	\item The second layer demands a {\bf semantically secure encryption scheme} to enable the tracing feature. The signer is constrained to encrypt its identifying information and to send the ciphertext as part of the group signature, so that the tracing authority can decrypt if and when necessary.
	This ingredient is also available in code-based cryptography, thanks to various \textsf{CPA}-secure and \textsf{CCA}-secure variants of the McEliece~\cite{McE78} and the Niederreiter~\cite{Nie86} cryptosystems available in, \eg, \cite{NIKM08,DDMN12,Persichetti12}, and \cite{MathewVVR12}.
	\item The third layer requires a {\bf zero-knowledge (\textsf{ZK}) protocol} that connects the previous two layers. This is essentially the bottleneck in realizing secure code-based group signatures. Specifically, the protocol should demonstrate that a given signature is generated by a certain certified group user who honestly encrypts its identifying information. Constructing such a protocol is quite challenging. There have been \textsf{ZK} protocols involving the CFS and Stern's signatures, which yield identity-based identification schemes in~\cite{CGG07,AlaouiCM11}, and \cite{YangTMSW14} and threshold ring signatures in~\cite{MCG08} and \cite{MelchorCGL11}. There have also been \textsf{ZK} proofs of plaintext knowledge for the McEliece and the Niederreiter cryptosystems~\cite{HuMT13}. Yet we are unaware of any efficient \textsf{ZK} protocol that \emph{simultaneously} deals with both code-based signature and encryption schemes in the above sense.
\end{enumerate}

Designing provably secure group signature schemes has been a long-standing open question in code-based cryptography, as was also discussed in~\cite{CM10}.

\subsection{Our Contributions}\label{subsec:Our-contributions}
This work introduces two group signature schemes which are provably secure under code-based assumptions. Specifically, our basic scheme achieves the CPA-anonymity~\cite{BBS04} and the traceability requirements in~\cite{BMW03} in the random oracle model. We assume the hardness of the McEliece problem, the Learning Parity with Noise problem, and a variant of the Syndrome Decoding problem. The basic scheme is then extended to achieve anonymity in the strongest sense~\cite{BMW03}, \ie, CCA-anonymity, for which the adversary is allowed to adaptively query for the opening of group signatures. Our two schemes are the first of their respective classes. 

\medskip

\noindent
{\bf Contributions to Code-Based Cryptography.} By introducing provably secure code-based group signature schemes, we solve the open problem discussed earlier. Along the way, we introduce the following two new techniques for code-based cryptography, which might be of independent interest.

\begin{enumerate}
	\item We design a \textsf{ZK} protocol for the randomized McEliece encryption scheme. The protocol allows the prover to convince the verifier that a given ciphertext is well-formed and that the hidden plaintext satisfies an additional condition. Such \emph{verifiable encryption protocols} are useful, not only in constructing group signatures, but also in much broader contexts~\cite{CamenischS03}. It is worth noting that, prior to our work, verifiable encryption protocols for code-based cryptosystems only exist in a very basic form where the plaintext is publicly given~\cite{HuMT13}, restricting their applications.
	
	\item In our security proof of the traceability property, to obtain a reduction from the hardness of the Syndrome Decoding (\textsf{SD}) problem, we come up with an approach that, to the best of our knowledge, has not been considered in the literature before. Let us recall the (average-case) \textsf{SD} problem with parameters $m, r, \omega$. Given a \emph{uniformly random} matrix $\widetilde{\mathbf{H}} \in \F_2^{r \times m}$ and a \emph{uniformly random} syndrome $\tilde{\mathbf{y}} \in \F_2^r$, the problem asks to find a vector $\mathbf{s} \in \F_2^m$ of Hamming weight $\omega$, denoted by $\mathbf{s} \in \mathsf{B}(m, \omega)$, such that $\widetilde{\mathbf{H}}\cdot\mathbf{s}^\top = \tilde{\mathbf{y}}^\top$.
	In our scheme, the key generation algorithm produces a public key that contains a matrix $\mathbf{H}\in \F_2^{r \times m}$ and syndromes $\mathbf{y}_j \in \F_2^r$, while users are given secret keys of the form $\mathbf{s}_j \in \mathsf{B}(m, \omega)$ such that $\mathbf{H}\cdot \mathbf{s}_j^\top = \mathbf{y}_j^\top$.
	In the security proof, since we would like to embed an \textsf{SD} challenge instance $(\widetilde{\mathbf{H}}, \tilde{\mathbf{y}})$ into the public key without being noticed, except with negligible probability, by the adversary, we have to require that $\mathbf{H}$ and the $\mathbf{y}_j$'s produced by the key generation are indistinguishable from uniform.
	
	One method to generate these keys is to employ the ``hash-and-sign'' technique from the CFS signature~\cite{CFS01}. Unfortunately, while the syndromes $\mathbf{y}_j$'s could be made uniformly random, as the outputs of the random oracle, the assumption that the CFS matrix $\mathbf{H}$ is \emph{computationally} close to uniform for practical parameters is invalidated by the distinguishing attacks from~\cite{FGOPT13}.
	
	Another method, pioneered by Stern~\cite{Ste96}, is to pick $\mathbf{H}$ and the $\mathbf{s}_j$'s uniformly at random. The corresponding syndromes $\mathbf{y}_j$'s could be made \emph{computationally} close to uniform if the parameters are set such that $\omega$ is slightly smaller than the value $\omega_0$ given by the Gilbert-Varshamov bound, \ie, $\omega_0$ such that $\binom{m}{\omega_0} \approx 2^r$. In such a case, the function $f_{\mathbf{H}}(\mathbf{s}_j) = \mathbf{H}\cdot \mathbf{s}_j^\top$ acts as a pseudorandom generator~\cite{FischerS96}. However, for these parameters, it is not guaranteed with high probability that a uniformly random \textsf{SD} instance $(\widetilde{\mathbf{H}}, \tilde{\mathbf{y}})$ has solutions, which would affect the success probability of the reduction algorithm.
	
	Our work considers the case when $\omega$ is moderately larger than $\omega_0$, so that two conditions hold. First, the uniform distribution over the set $\mathsf{B}(m,\omega)$ has sufficient min-entropy to apply the Left-over Hash Lemma from~\cite{GKPV10}. Second, the \textsf{SD} problem with parameters $(m, r, \omega)$ admits solutions with high probability, yet remains intractable against the best known attacks from~\cite{FS09} and~\cite{BeckerJMM12}. Note that the variant of the \textsf{SD} problem considered in this work are not widely believed to be the hardest one~\cite{Ste96,Meurer-thesis}, but suitable parameters can be chosen such that the best known attacks run in exponential time. Further treatment on how to decide on the parameters will be given in Section~\ref{sec:implement}. This approach gives us a new method to generate uniformly random vectors $\mathbf{s}_j \in \mathsf{B}(m,\omega)$ and a matrix $\mathbf{H} \in \F_2^{r \times m}$ so that the syndromes $\mathbf{y}_j$'s corresponding to the $\mathbf{s}_j$'s are \emph{statistically} close to uniform. The approach, which somewhat resembles the technique used in~\cite{GPV08} for the Inhomogeneous Small Integer Solution problem, is helpful in our security proof and, generally, in designing formal security reductions from the \textsf{SD} problem.
\end{enumerate}

\medskip

\noindent
{\bf Contributions to Group Signatures from Quantum-Resistant Assumptions.} {Although we have not obtained security proofs in the quantum random oracle model (QROM), our constructions provide the first non-lattice-based alternatives to provably secure group signatures from quantum-resistant assumptions.} Our schemes feature public key and signature sizes linear in the number of group users $N$, which are \emph{asymptotically} not as efficient as the previously published lattice-based counterparts~\cite{LLLS13-Asiacrypt,LLNW14-PKC,LNW15,NZZ15}. However, when instantiated with practical parameters, our schemes behave much more efficiently than the scheme proposed in~\cite{NZZ15}. The latter is arguably the current most efficient lattice-based group signature in the asymptotic sense. Indeed, our estimation shows that our basic scheme, which achieves the \textsf{CPA}-anonymity notion, gives public key and signature sizes that are $2,300$ times and $540$ times smaller, respectively, for an average-size group of $N=2^8$ users. As $N$ grows, the advantage lessens, but our basic scheme remains more efficient even for a huge group of $N=2^{24}$ users, a number which is roughly comparable to the current population of the Netherlands. 
Our extended scheme, which achieves the strongest anonymity notion, \ie, \textsf{CCA}-anonymity, introduces only a small overhead of about $434$ KB  and $177$~KB in public key and signature sizes, respectively, compared to the basic scheme. 
Table~\ref{Table:Comparison} gives the details of our estimation. The parameters for our schemes are set as in Section~\ref{sec:implement}. For the scheme in~\cite{NZZ15}, we choose the commonly used lattice dimension $n=2^8$ and set the parameters $m=2^9\times 150$ and $q=2^{150}$ to satisfy the requirements given in~\cite[Section~5.1]{NZZ15}. While our basic scheme and the scheme in~\cite{NZZ15} achieve the \textsf{CPA}-anonymity notion~\cite{BBS04}, our extended scheme achieves the \textsf{CCA}-anonymity notion~\cite{BMW03}. All schemes have soundness error $2^{-80}$.

We give actual implementation results for our proposed schemes to support their claim of feasibility. In our implementations, as presented later in Section~\ref{sec:implement}, the actual signature sizes can be reduced thanks to an additional technique. Our schemes are {the first group signature from quantum-resistant assumptions} that comes supported with an actual deployment analysis. The results, while not yielding a truly practical scheme, certainly help in bringing {this new class of group signatures} closer to practice.

\begin{table*}[!htbp]
\begin{center}
\renewcommand{\arraystretch}{1.3}
\caption{Efficiency comparison between our schemes and~\cite{NZZ15}.}\label{Table:Comparison}
\begin{tabular}{|c|c|c|rr|rr|}
\hline
& Anonymity & $N$ & \multicolumn{2}{c|}{Public Key Size} & \multicolumn{2}{c|}{Signature Size}\\
\hline
\hline
\multirow{6}{*}{Ours} & \multirow{3}{*}{\textsf{CPA}} & $2^{8}$ & $5.13\times 10^{6}$~bits & (\,642\,~KB)~& $8.57\times 10^{6}$~bits & (1.07~MB)~\\
\cline{3-7}
&& $2^{16}$ & $4.10\times 10^{7}$~bits & (5.13~MB)~& $1.77\times 10^{7}$~bits & (2.21~MB)~\\
\cline{3-7}
&& $2^{24}$ & $9.23\times 10^{9}$~bits & (1.16\,~GB)~& $2.36\times 10^{9}$~bits & (294\,~MB)~\\
\cline{2-7}\cline{2-7}
 & \multirow{3}{*}{\textsf{CCA}} & $2^{8}$ & $8.60\times 10^{6}$~bits & (1.08~MB)~& $9.99\times 10^{6}$~bits & (1.25~MB)~\\
\cline{3-7}
&& $2^{16}$ & $4.45\times 10^{7}$~bits & (5.56~MB)~& $1.91\times 10^{7}$~bits & (2.39~MB)~\\
\cline{3-7}
&& $2^{24}$ & $9.23\times 10^{9}$~bits & (1.16\,~GB)~& $2.36\times 10^{9}$~bits & (294\,~MB)~\\
\hline
\hline
\cite{NZZ15} & \textsf{CPA} &  $\leq 2^{24}$ & $1.18\times 10^{10}$~bits & (1.48\,~GB)~& $4.63\times 10^{9}$~bits & (579\,~MB)~\\
\hline
\end{tabular}
\end{center}
\end{table*}

\subsection{Overview of Our Techniques}
Let $m, r, \omega, \nmc, \kmc, \tmc$ and $\ell$ be positive integers. We consider a group of size $N = 2^\ell$, where each user is indexed by an integer $j \in [0, N-1]$.
The secret signing key of user $j$ is a vector $\mathbf{s}_j$ chosen uniformly at random from the set $\mathsf{B}(m,\omega)$. A uniformly random matrix $\mathbf{H} \in \F_2^{r \times m}$ and $N$ syndromes $\y_0, \ldots, \y_{N-1} \in \F_2^r$, such that $\mathbf{H}\cdot\mathbf{s}_j^\top = \mathbf{y}_j^\top$, for all $j$, are made public. Let us now explain the development of the three ingredients used in our basic scheme.

\smallskip

\noindent
{\bf The Signature Layer.}
User $j$ can run Stern's \textsf{ZK} protocol~\cite{Ste96} to prove the possession of a vector $\mathbf{s} \in \mathsf{B}(m,\omega)$ such that $\mathbf{H}\cdot\mathbf{s}^\top = \mathbf{y}_j^\top$. The constraint $\mathbf{s} \in \mathsf{B}(m,\omega)$ is proved in \textsf{ZK} by randomly permuting the entries of $\mathbf{s}$ and showing that the permuted vector belongs to $\mathsf{B}(m,\omega)$.
The protocol is then transformed into a Fiat-Shamir signature~\cite{FS86}. However, such a signature is publicly verifiable only if the index $j$ is given to the verifier.

The user can further hide its index $j$ to achieve unconditional anonymity among all $N$ users. This, incidentally, yields a \emph{ring signature}~\cite{RivestST01} on the way, \`{a} la~\cite{BettaiebS13}. Let $\mathbf{A} = \big[\y_0^\top | \cdots | \y_j^\top| \cdots | \y_{N\hspace*{-1.5pt}-\hspace*{-1.5pt}1}^\top \big] \in \F_2^{r \times N}$. Let $\mathbf{x} = \delta_j^N$ be the $N$-dimensional unit vector with entry $1$ at the $j$-th position and $0$ elsewhere. Observe that $\mathbf{A}\cdot \mathbf{x}^\top = \y_j^\top$, and thus, the equation $\mathbf{H}\cdot \mathbf{s}^\top = \y_j^\top$ can be written as
\begin{equation}\label{eqn:intro-signing-layer}
\mathbf{H}\cdot \mathbf{s}^\top \oplus \mathbf{A}\cdot \mathbf{x}^\top = \mathbf{0},
\end{equation}
where $\oplus$ denotes addition modulo $2$. Stern's framework allows the user to prove in \textsf{ZK} the possession of $(\mathbf{s}, \mathbf{x})$ satisfying this equation, where the condition $\mathbf{x} = \delta_j^N$ can be justified using a random permutation.

\smallskip

\noindent
{\bf The Encryption Layer.} To enable the tracing capability of the scheme, we let user $j$ encrypt the binary representation of $j$ via the randomized McEliece encryption scheme~\cite{NIKM08}. Specifically, we represent $j$ as $\itob(j) = (j_0, \ldots, j_{\ell-1}) \in \{0,1\}^\ell$, where $\sum_{i=0}^{\ell-1}j_i 2^{\ell-1-i} = j$. Given a public encrypting key $\mathbf{G} \in \F_2^{k \times n}$, a ciphertext of $\itob(j)$ is of the form 
\begin{equation}\label{eqn:intro-encryption-layer}
\mathbf{c} = \big(\hspace*{1.5pt}\mathbf{u}\hspace*{1.5pt} \|\hspace*{1.5pt}\itob(j)\hspace*{1.5pt}\big)\cdot\Gmc \oplus \mathbf{e} \in \F_2^{\nmc},
\end{equation}
where $(\mathbf{u}, \mathbf{e})$ is the encryption randomness, with $\mathbf{u} \in \F_2^{k - \ell}$, and $\mathbf{e} \in \mathsf{B}(n, t)$, \ie, $\mathbf{e}$ is a vector of weight $t$ in $\F_2^n$.

\smallskip

\noindent
{\bf Connecting the Signature and Encryption Layers.} User $j$ must demonstrate that it does not cheat, \eg, by encrypting some string that does not point to $j$, without revealing $j$. Thus, we need a \textsf{ZK} protocol that allows the user to prove that the vector $\mathbf{x} = \delta_j^N$ used in~(\ref{eqn:intro-signing-layer}) and the plaintext hidden in~(\ref{eqn:intro-encryption-layer}) both correspond to the same secret $j \in [0,N-1]$. The crucial challenge is to establish a connection, which must be verifiable in \textsf{ZK}, between the ``index representation'' $\delta_j^N$ and the binary representation $\itob(j)$. We show how to handle this challenge well.

Instead of working with $\itob(j) = (j_0, \ldots, j_{\ell-1})$, let us consider an extension of $\itob(j)$, defined as 
\[
\Encode(j) = (1-j_0, j_0, \ldots, 1-j_i, j_i, \ldots, 1- j_{\ell-1}, j_{\ell-1}) \in \F_2^{2\ell}.
\]
We then suitably insert $\ell$ zero-rows into $\mathbf{G}$ to obtain $\widehat{\mathbf{G}} \in \F_2^{(k + \ell) \times n}$ such that $\big(\hspace*{1.5pt}\mathbf{u}\hspace*{1.5pt}\|\hspace*{1.5pt}\Encode(j)\hspace*{1.5pt}\big)\cdot\widehat{\Gmc}= \big(\hspace*{1.5pt}\mathbf{u}\hspace*{1.5pt}\|\hspace*{1.5pt}\itob(j)\hspace*{1.5pt}\big)\cdot\Gmc$. Letting $\mathbf{f} = \Encode(j)$, we rewrite (\ref{eqn:intro-encryption-layer}) as
\begin{equation}\label{eqn:intro-encryption-layer-2}
\mathbf{c} = \big(\hspace*{1.5pt}\mathbf{u}\hspace*{1.5pt}\|\hspace*{1.5pt}\mathbf{f}\hspace*{1.5pt}\big)\cdot\widehat{\Gmc} \oplus \mathbf{e} \in \F_2^{\nmc}.
\end{equation}
Now, let $\btoi: \{0,1\}^\ell \rightarrow [0,N-1]$ be the inverse function of $\itob(\cdot)$. For every $\mathbf{b} \in \{0,1\}^\ell$, we carefully design two classes of permutations $T_{\mathbf{b}}: \F_2^N \rightarrow \F_2^N$ and $T'_{\mathbf{b}}: \F_2^{2\ell} \rightarrow \F_2^{2\ell}$, such that, for any $j \in [0,N-1]$, 
\[
\mathbf{x} = \delta_j^N \iff T_{\mathbf{b}}(\mathbf{x}) = \delta^N_{\btoi(\itob(j)\oplus \mathbf{b})} \mbox{ and }
\mathbf{f} = \Encode(j) \iff  T'_{\mathbf{b}}(\mathbf{f})=\Encode(\btoi(\itob(j)\oplus \mathbf{b})).
\]
Given the equivalences, the protocol's user samples a uniformly random vector $\mathbf{b} \in \{0,1\}^\ell$ and sends $\mathbf{b}_1 = \itob(j)\oplus \mathbf{b}$. The verifier, seeing that 
\[
T_{\mathbf{b}}(\mathbf{x}) = \delta^N_{\btoi(\mathbf{b}_1)} \mbox{ and } T'_{\mathbf{b}}(\mathbf{f})=\Encode(\btoi(\mathbf{b}_1)),
\]
should be convinced that $\mathbf{x}$ and $\mathbf{f}$ correspond to the same $j \in [0,N-1]$, yet the value of $j$ is completely hidden from its view since $\mathbf{b}$ acts essentially as a one-time pad.

The technique extending $\itob(j)$ into $\Encode(j)$ and then permuting $\Encode(j)$ in a ``one-time pad'' fashion is inspired by a method originally proposed by Langlois \emph{et al.} in \cite{LLNW14-PKC} in a seemingly unrelated context. There, the goal is to prove that the message being signed under the Bonsai tree signature~\cite{CHKP10} is of the form $\itob(j)$, for some $j \in [0,N-1]$. Here, we adapt and develop their method to \emph{simultaneously} prove two facts. First, the plaintext being encrypted under the randomized McEliece encryption is of the form $\itob(j)$. Second, the unit vector $\mathbf{x} = \delta_j^N$ is used in the signature layer.

By embedding the above technique into Stern's framework, we obtain an interactive \textsf{ZK} argument system, in which, given the public input $(\mathbf{H}, \mathbf{A}, \mathbf{G})$, the user is able to prove the possession of a secret tuple $(j, \mathbf{s}, \mathbf{x}, \mathbf{u}, \mathbf{f}, {\mathbf{e}})$ satisfying~(\ref{eqn:intro-signing-layer}) and~(\ref{eqn:intro-encryption-layer-2}). The protocol is repeated many times to achieve negligible soundness error, and then made non-interactive, resulting in a non-interactive \textsf{ZK} argument of knowledge $\Pi$. The final group signature is of the form $(\mathbf{c}, \Pi)$, where $\mathbf{c}$ is the ciphertext. In the random oracle model, the anonymity of the scheme relies on the zero-knowledge property of $\Pi$ and the \textsf{CPA}-security of the randomized McEliece encryption scheme, while its traceability is based on the hardness of the variant of the \textsf{SD} problem discussed earlier.

\smallskip

\noindent
{\bf Achieving CCA-Anonymity.} Our basic group signature scheme makes use of the randomized McEliece encryption scheme that achieves only \textsf{CPA}-security. Hence, it only satisfies \textsf{CPA}-anonymity for which the adversary is not granted access to the signature opening oracle. To achieve the strongest notion of anonymity put forward in~\cite{BMW03}, \ie, \textsf{CCA}-anonymity, we would need a \textsf{CCA2}-secure encryption scheme so that we can respond to adaptive opening queries from the adversary by invoking the decryption oracle associated with the encryption mechanism. There are a number of known \textsf{CCA2}-secure code-based encryption schemes, \eg, \cite{DDMN12,Persichetti12,MathewVVR12}. They are, however, either too inefficient, say with ciphertext size quadratic in the security parameter, or incompatible with zero-knowledge protocols for proving the well-formedness of ciphertexts. Hence, they are unsuitable for our purpose. Instead, we exploit the fact that our verifiable encryption protocol for the randomized McEliece scheme is a simulation-sound \textsf{ZK} argument of knowledge. We then upgrade the encryption system further to a \textsf{CCA2}-secure one via the Naor-Yung twin-encryption paradigm~\cite{NY90}. The protocol operates in Stern's framework and satisfies the ``quasi-unique responses'' property in~\cite{BCKLN14}, deriving simulation-soundness from soundness. This fact was recently exploited by several group signature schemes, such as \cite{LLNW16} and \cite{LNWX18}, which are based on Stern-like protocols.

Specifically, we will work with two public keys $\mathbf{G}^{(1)}$ and $\mathbf{G}^{(2)}$ of the randomized McEliece encryption scheme. The user now encrypts $\itob(j)$ under each of the keys to obtain ciphertexts $\mathbf{c}^{(1)}$ and $\mathbf{c}^{(2)}$, respectively, and extend the verifiable encryption protocol discussed above to prove that these ciphertexts are well-formed and correspond to the same plaintext $\itob(j)$, which is the binary representation of the user's index $j$. This extension is quite smooth, since the same techniques for handling ciphertext $\mathbf{c}$ can be used to handle $\mathbf{c}^{(1)}$ and $\mathbf{c}^{(2)}$. In the proof of \textsf{CCA}-anonymity, we then employ the strategy of~\cite{Sahai99} that makes use of the \textsf{CPA}-security of the underlying encryption scheme and the zero-knowledge, soundness and simulation-soundness of the resulting non-interactive argument. In terms of efficiency, our \textsf{CCA}-anonymous construction only has a small and reasonable overhead compared to its \textsf{CPA}-anonymous version, with one more McEliece encrypting matrix in the group public key and one more ciphertext equipped with its supporting \textsf{ZK} sub-protocol in the group signature. 

\subsection{Related Works}
The present paper is the full extension of our earlier work~\cite{ELLNW15}, which was published in the proceedings of ASIACRYPT 2015. Achieving \textsf{CCA}-anonymity for code-based group signatures was raised as an open question in~\cite{ELLNW15}. We are able to fully address the problem in this work. 

In a work concurrent to and independent of~\cite{ELLNW15}, Alam{\'{e}}lou \emph{et al.} also proposed a code-based group signature scheme in~\cite{ABCG15} and, later on, in~\cite{ABCG17}. Their scheme considers the setting of dynamic groups. It does not use any encryption mechanism to enable traceability. Instead, the authors rely on a modified version of Stern's protocol that allows the opening authority to test whether each protocol execution is generated using a secret key of a given user. 
Unfortunately, such approach does not yield a secure group signature. Recall that Stern's protocol admits a soundness error of $2/3$ in each execution. It has to be repeated $\kappa = \omega(\log \lambda)$ times, where $\lambda$ is the security parameter, to make the error negligibly small. Then, a valid signature is generated by an honest user $j$ if and only if the tests for \emph{all} $\kappa$ executions of the protocol yield the \emph{same} user $j$. Unfortunately, the testing mechanism used in their scheme fails to capture this crucial point. When running through protocol execution numbers $1, 2, \ldots, \kappa$, it stops and outputs user $j$ when it sees the \emph{first} execution that points to $j$. This shortcoming opens a room for cheating users to break the traceability and non-frameability of the scheme. Specifically, a cheating user $j'$, who wants to mislead the opening result to an innocent user $j$, can simulate the first several protocol executions. The simulation can be done with noticeable probability using the transcript simulator associated with the protocol, because each execution admits a soundness error of $2/3$. If the opening algorithm is run, it would return $j$ with noticeable probability. The remaining protocol executions are done faithfully with secret key for user $j'$. Thus, the construction in~\cite{ABCG15} and~\cite{ABCG17} is not secure. We note that a very similar testing mechanism for Stern-like protocols was used in~\cite{LLNW14-PKC} to avoid the use of encryption in their group signature. This had been broken. In~\cite{LNRW18}, which is the corrected version of~\cite{LLNW14-PKC}, the authors eventually had to rely on an encryption-like mechanism to enable traceability.

In a very recent work, Nguyen \emph{et al.}~\cite{NTWZ19} proposed a number of new code-based privacy-preserving cryptographic constructions, including the first code-based group signature scheme with logarithmic signature size – which resolves an interesting question we left open in~\cite{ELLNW15}. In their scheme, group users are associated with leaves in a code-based Merkle tree supported by a zero-knowledge argument of tree inclusion, which has communication cost linear in the tree depth (and hence, logarithmic in the number of users). Although their scheme achieves better asymptotic efficiency than ours, it yields signature size larger than $2.5$~MB even for small groups. In particular, for groups of size up to $2^{16}$, the signatures are longer than those produced by our \textsf{CPA}-anonymous and \textsf{CCA}-anonymous schemes.

Subsequent to the publication of~\cite{ELLNW15}, a number of lattice-based group signatures have been proposed, bringing {group signatures from quantum-resistant assumptions} much closer to practice. Examples include the works done in~\cite{LLNW16,LLMNW16,LNWX18,BCN18}, and \cite{PLS18}. We believe that this interesting research direction will continue to attract attention from the community. The hope is that some provably secure and truly practical schemes can be realized in the near future.

\section{Preliminaries}\label{Sec:Prelim}
{\sc Notations.}
Let $\lambda$ be the security parameter and $\mathsf{negl}(\lambda)$ denote a negligible function in $\lambda$. We use $a\lr A$ if $a$ is chosen uniformly at random from the finite set $A$.
The symmetric group of all permutations of $k$ elements is denoted by $\mathsf{S}_k$. Bold capital letters, \eg, $\mathbf{A}$, denote matrices. Bold lowercase letters, \eg, $\mathbf {x}$, denote row vectors. We use ${\text{\bf x}}^{\top}$ to denote the transpose of ${\text{\bf x}}$ and $wt(\mathbf{x})$ to denote the (Hamming) weight of $\mathbf{x}$. We denote by $\mathsf{B}(m,\omega)$ the set of all vectors $\mathbf{x} \in \F_2^m$ such that $wt(\mathbf{x}) = \omega$.
Throughout the paper, we define a function $\itob$ which takes a non-negative integer $a$  as an input, and outputs the binary representation $(a_{0}, \cdots, a_{\ell-1})\in\{0,1\}^{\ell}$ of $a$ such that $a=\sum_{i=0}^{\ell-1}a_i2^{\ell-1-i}$, and a function $\btoi$ which takes as an input the binary representation $(a_{0}, \cdots, a_{\ell-1})\in\{0,1\}^{\ell}$ of $a$, and outputs $a$.
All logarithms are in base $2$.

\subsection{Background on Code-Based Cryptography}

We first recall the Syndrome Decoding problem. It is well-known to be \textsf{NP}-complete~\cite{Berlekamp78}, and is widely believed to be intractable in the average case for appropriate choice of parameters~\cite{Ste96,Meurer-thesis}.

\begin{definition}[The Syndrome Decoding problem]
The $\mathsf{SD}(m, r, \omega)$ problem asks, given a uniformly random matrix $\mathbf{H} \in \F_2^{r \times m}$ and a uniformly random syndrome $\mathbf{y}\in \F_2^r$, for a vector $\mathbf{s} \in \mathsf{B}(m, \omega)$ such that $\mathbf{H}\cdot\mathbf{s}^\top = \mathbf{y}^\top$. When $m = m(\lambda)$, $r= r(\lambda)$ and $\omega= \omega(\lambda)$, we say that the $\mathsf{SD}(m, r, \omega)$ problem is hard if the success probability of any $\mathrm{PPT}$ algorithm in solving the problem is at most $\mathsf{negl}(\lambda)$.
\end{definition}

In our security reduction, the following variant of the Left-over Hash Lemma for matrix multiplication over $\F_2$ is used.
\begin{lemma}[Left-over Hash Lemma, adapted from~\cite{GKPV10}] \label{lemma:Leftover-Hash-Lemma}
Let $D$ be a distribution over $\F_2^m$ with min-entropy $e$. For $\epsilon > 0$ and $r \leq e - 2\log(1/\epsilon) -\mathcal{O}(1)$, the statistical distance between the distribution of $(\mathbf{H}, \mathbf{H}\cdot\mathbf{s}^\top)$, where $\mathbf{H} \xleftarrow{\$} \F_2^{r \times m}$ and $\mathbf{s} \in \F_2^m$ is drawn from distribution $D$, and the uniform distribution over $\F_2^{r \times m} \times \F_2^r$ is at most $\epsilon$.
	
In particular, if $\omega <m$ is an integer such that $r \leq \log\binom{m}{\omega} - 2\lambda - \mathcal{O}(1)$ and $D$ is the uniform distribution over $\mathsf{B}(m,\omega)$, \ie, $D$ has min-entropy $\log \binom{m}{\omega}$, then the statistical distance between the distribution of $(\mathbf{H}, \mathbf{H}\cdot\mathbf{s}^\top)$ and the uniform distribution over $\F_2^{r \times m} \times \F_2^r$ is at most $2^{-\lambda}$.
\end{lemma}

\smallskip

\noindent
{\bf The Randomized McEliece Encryption Scheme.}
We employ the following randomized variant, suggested in~\cite{NIKM08}, of the McEliece encryption scheme~\cite{McE78}, where a uniformly random vector is concatenated to the plaintext.
\begin{enumerate}
	\item $\mcsetup(1^\lambda)$: Select the parameters $\nmc = \nmc(\lambda), \kmc = \kmc(\lambda), \tmc = \tmc(\lambda)$ for a binary $[\nmc, \kmc, 2\tmc+1]$ Goppa code. 
	Choose integers $k_1$ and $k_2$ such that $\kmc=k_1+k_2$. Set the plaintext space as $\F_{2}^{k_2}$.
	\item $\mckeygen(\nmc, \kmc, \tmc)$: Perform the following steps.
	\begin{enumerate}
		\item Produce a generator matrix $\Gmc' \in \F_2^{\kmc\times\nmc}$ of a randomly selected $[\nmc, \kmc, 2\tmc+1]$ Goppa code. 
		Choose a random invertible matrix $\Smc \in \F_2^{\kmc\times\kmc}$ and a random permutation matrix $\Pmc \in \F_2^{\nmc\times\nmc}$. Let $\Gmc=\Smc\Gmc'\Pmc \in \F_2^{\kmc \times \nmc}$.
		\item Output encrypting key $\pkmc=\Gmc$ and decrypting key $\skmc=(\Smc, \Gmc', \Pmc)$.	
	\end{enumerate}
	\item $\mcenc(\pkmc, \mathbf{m})$: To encrypt a message $\mathbf{m}\in\F_2^{k_2}$,
	sample $\mathbf{u} \xleftarrow{\$}\F_2^{k_1}$ and $\mathbf{e} \lr \mathsf{B}(\nmc, \tmc)$,
	then output the ciphertext $\mathbf{c}=(\mathbf{u}\|\mathbf{m})\cdot\Gmc \oplus \mathbf{e} \in \F_2^{\nmc}$.
	\item $\mcdec(\skmc, \mathbf{c})$: Perform the following steps.
	\begin{enumerate}
		\item Compute $\mathbf{c}\cdot\Pmc^{-1}=((\mathbf{u}\|\mathbf{m})\cdot\Gmc \oplus \mathbf{e})\cdot\Pmc^{-1}$ and then $\mathbf{m}'\cdot\Smc = \decode_{\Gmc'}(\mathbf{c}\cdot\Pmc^{-1})$ where $\decode$ is an error-correcting algorithm with respect to $\Gmc'$. If $\decode$ fails, then return $\bot$.
		\item Compute $\mathbf{m}'=(\mathbf{m}'\Smc)\cdot\Smc^{-1}$, parse $\mathbf{m}'=(\mathbf{u}\|\mathbf{m})$, where $\mathbf{u}\in\F_2^{k_1}$ and $\mathbf{m}\in\F_2^{k_2}$, and return $\mathbf{m}$.
	\end{enumerate}	
\end{enumerate}
The scheme described above is \textsf{CPA}-secure in the standard model, assuming the hardness of the $\mathsf{DMcE}(\nmc, \kmc, \tmc)$ problem and the $\mathsf{DLPN}(\kmc_1, \nmc, \mathsf{B}(\nmc, \tmc))$ problem~\cite{NIKM08,Doettling-thesis}. We now recall these two problems.

\begin{definition}[The Decisional McEliece problem] 
The $\mathsf{DMcE}(\nmc, \kmc, \tmc)$ problem is to distinguish if a given $\mathbf{G} \in \F_2^{\kmc \times \nmc}$ is a uniformly random matrix over $\F_2^{\kmc \times \nmc}$ or is generated by $\mckeygen(\nmc, \kmc, \tmc)$ above. When $\nmc= \nmc(\lambda), \kmc= \kmc(\lambda), \tmc= \tmc(\lambda)$, we say that the $\mathsf{DMcE}(\nmc, \kmc, \tmc)$ problem is hard if the success probability of any $\mathrm{PPT}$ distinguisher is at most $1/2 + \mathsf{negl}(\lambda)$.
\end{definition}

\begin{definition}[The Decisional Learning Parity with (fixed-weight) 
Noise problem] The $\mathsf{DLPN}(\kmc, \nmc, \mathsf{B}(\nmc, \tmc))$ problem, given a pair $(\mathbf{A}, \mathbf{v}) \in \F_2^{k \times n} \times \F_2^n$, is to distinguish whether $(\mathbf{A}, \mathbf{v})$ is a uniformly random pair over $\F_2^{k \times n} \times \F_2^n$ or  is obtained by choosing $\mathbf{A} \lr \F_2^{k \times n}$, $\mathbf{u} \lr \F_2^k$, $\mathbf{e} \lr \mathsf{B}(\nmc, \tmc)$ and outputting $(\mathbf{A}, \mathbf{u}\cdot \mathbf{A} \oplus \mathbf{e})$. When $\kmc = \kmc(\lambda), \nmc=\nmc(\lambda), \tmc=\tmc(\lambda)$, we say that the $\mathsf{DLPN}(\kmc, \nmc, \mathsf{B}(\nmc, \tmc))$ problem is hard, if the success probability of any $\mathrm{PPT}$ distinguisher is at most $1/2 + \mathsf{negl}(\lambda)$.
\end{definition}

\subsection{Group Signatures}
We follow the definition of group signatures provided in~\cite{BMW03} for the case of static groups.
\begin{definition}
A {\it group signature} 
\[
\mathcal{GS}= \textsf{(KeyGen, Sign, Verify, Open)}
\]
is a tuple of the following four polynomial-time algorithms.
\begin{enumerate}
\item \textsf{KeyGen}: This randomized algorithm takes as input $(1^\lambda, 1^{N})$, where $N \in \mathbb{N}$ is the number of group users, and outputs $\mathsf{(gpk, gmsk, gsk)}$, where $\mathsf{gpk}$ is the group public key, $\mathsf{gmsk}$ is the group manager's secret key, and $\mathsf{gsk}= \{\mathsf{gsk}[j]\}_{j \in [0,N-1]}$ with $\mathsf{gsk}[j]$ being the secret key for the group user of index $j$.
\item \textsf{Sign}: This randomized algorithm takes as input a secret signing key $\mathsf{gsk}[j]$ for some $j \in[0, N-1]$ and a message $M$ and returns a group signature $\Sigma$ on $M$.
\item \textsf{Verify}: This deterministic algorithm takes as input the group public key $\mathsf{gpk}$, a message $M$, a signature $\Sigma$ on $M$. The output is either $1$ \textsf{(Accept)} or $0$ \textsf{(Reject)}.
\item \textsf{Open}: This deterministic algorithm takes as input the group manager's secret key $\mathsf{gmsk}$, a message $M$, a signature $\Sigma$ on $M$. It outputs either an index $j \in[0,N-1]$, which is associated with a particular user, or $\bot$, indicating failure.
\end{enumerate}
\end{definition}

\smallskip

A {\bf correct} group signature scheme requires that, for all $\lambda, N \in \mathbb{N}$, all $\mathsf{(gpk, gmsk, gsk)}$ produced by $\textsf{KeyGen}(1^\lambda, 1^N)$, all $j \in[0,N-1]$, and all messages $M \in \{0,1\}^*$, we have 
\[
\textsf{Verify}\big(\mathsf{gpk}, M, 
\textsf{Sign}(\mathsf{gsk}[j], M)\big)=1 \mbox{ and } 
\textsf{Open}\big(\mathsf{gmsk}, M, \textsf{Sign}(\mathsf{gsk}[j], M)\big)=j.
\]

A {\bf secure} group signature scheme must meet two security conditions.
\begin{enumerate}
	\item \emph{Traceability}: all signatures, even those produced by a coalition of group users and the group manager, can be traced back to a member of the coalition.
	\item \emph{Anonymity}: signatures generated by two distinct group users are computationally indistinguishable to an adversary who knows all of the user secret keys. In Bellare \emph{et al.}'s model~\cite{BMW03}, the anonymity adversary is granted access to an opening oracle (\textsf{CCA}-anonymity). A relaxed notion, where the adversary cannot query the opening oracle (\textsf{CPA}-anonymity), was later proposed by Boneh \emph{et al.}~\cite{BBS04}. 
\end{enumerate}

We now give the formal definitions of \textsf{CPA}-anonymity, \textsf{CCA}-anonymity and traceability.

\begin{definition}
A group signature $\mathcal{GS}= \mathsf{(KeyGen, Sign, Verify, Open)}$ is $\mathsf{CPA}$-anonymous if, for all polynomial $N(\cdot)$ and any PPT adversaries $\mathcal{A}$, the advantage of $\mathcal{A}$ in the following experiment is negligible in $\lambda$.
\begin{enumerate}
	\item Run $(\mathsf{gpk}, \mathsf{gmsk}, \mathsf{gsk})\leftarrow \mathsf{KeyGen}(1^{\lambda}, 1^{N})$ and send $(\mathsf{gpk}, \mathsf{gsk})$ to $\mathcal{A}$.		
	\item $\mathcal{A}$ outputs two identities $j_0,j_1\in[0, N-1]$ with a message $M^*$. Choose a random bit $b$ and give $\Sigma^* \leftarrow \mathsf{Sign}(\mathsf{gsk}[j_b], M^*)$ to $\mathcal{A}$. Then, $\mathcal{A}$ outputs a bit $b'$.
\end{enumerate}
If $b'=b$, then $\mathcal{A}$ succeeds. The advantage of $\mathcal{A}$ is defined to be $\left|\Pr[\mathcal{A}~\text{succeeds}]-\dfrac{1}{2}\right|$.

A group signature $\mathcal{GS}= \mathsf{(KeyGen, Sign, Verify, Open)}$ is $\mathsf{CCA}$-anonymous if for all polynomial $N(\cdot)$ and any PPT adversaries $\mathcal{A}$, the advantage of $\mathcal{A}$ in the following experiment is negligible in $\lambda$.
\begin{enumerate}
	\item Run $(\mathsf{gpk}, \mathsf{gmsk}, \mathsf{gsk})\leftarrow \mathsf{KeyGen}(1^{\lambda}, 1^{N})$ and send $(\mathsf{gpk}, \mathsf{gsk})$ to $\mathcal{A}$.
		
	\item $\mathcal{A}$ can make queries to the opening oracle. On input a message $M$ and a signature $\Sigma$, the oracle returns $\mathsf{Open}(\mathsf{gmsk}, M, \Sigma)$ to $\mathcal{A}$. 
		
	\item $\mathcal{A}$ outputs two identities $j_0,j_1\in[0, N-1]$ with a message $M^*$. Choose a random bit $b$ and give $\Sigma^* \leftarrow \mathsf{Sign}(\mathsf{gsk}[j_b], M^*)$ to $\mathcal{A}$. 
		
	\item $\mathcal{A}$ can make further queries to the opening oracle, with the exception that it cannot query for the opening of $(M^*, \Sigma^*)$. 
		
	\item Finally, $\mathcal{A}$ outputs a bit $b'$.
\end{enumerate}
$\mathcal{A}$ succeeds if $b'=b$. The advantage of $\mathcal{A}$ is defined to $\left|\Pr[\mathcal{A}~\text{succeeds}]-\dfrac{1}{2}\right|$.

A group signature $\mathcal{GS}= \mathsf{(KeyGen, Sign, Verify, Open)}$ is traceable if for all polynomial $N(\cdot)$ and any PPT adversaries $\mathcal{A}$, the success probability of $\mathcal{A}$ in the following experiment is negligible in $\lambda$.
\begin{enumerate}
	\item Run $(\mathsf{gpk}, \mathsf{gmsk}, \mathsf{gsk})\leftarrow \mathsf{KeyGen}(1^{\lambda}, 1^{N})$ and send $(\mathsf{gpk}, \mathsf{gmsk})$ to $\mathcal{A}$.
		
	\item $\mathcal{A}$ may query the following oracles adaptively and in any order.
	\begin{enumerate}
		\item An $\mathcal{O}^\mathsf{Corrupt}$ oracle that on input $j\in[0, N-1]$, outputs $\mathsf{gsk}[j]$.
			
		\item An $\mathcal{O}^\mathsf{Sign}$ oracle that on input $j$, a message $M$, returns $\mathsf{Sign}(\mathsf{gsk}[j], M)$.
			
	\end{enumerate}
	Let $CU$ be the set of identities queried to $\mathcal{O}^\mathsf{Corrupt}$.
		
	\item Finally, $\mathcal{A}$ outputs a message $M^*$ and a signature $\Sigma^{*}$.	
\end{enumerate}
$\mathcal{A}$ succeeds if $\mathsf{Verify}(\mathsf{gpk}, M^*, \Sigma^*)=1$ and $\mathsf{Sign}(\mathsf{gsk}[j], M^*)$ was never queried for $j\notin CU$, yet $\mathsf{Open}(\mathsf{gmsk}, M^*, \Sigma^*)\notin CU$.
\end{definition}

\section{The Underlying Zero-Knowledge Argument Systems}\label{sec:nizk}
A statistical zero-knowledge argument system is an interactive protocol where the soundness property holds for \emph{computationally bounded} cheating provers, while the zero-knowledge property holds against \emph{any} cheating verifier. In this section we present statistical zero-knowledge argument systems which will serve as building blocks in our CPA-anonymous and CCA-anonymous group signature schemes in Section~\ref{sec:main-construction} and Section~\ref{sec:CCA}, respectively.

Before describing the protocols, we introduce several supporting notations and techniques. Let $\ell$ be a positive integer and $N=2^\ell$. For $\mathbf{x} = (x_0, x_1, \ldots, x_{N-1}) \in \F_2^N$ and for $j \in [0, N-1]$, we write $\mathbf{x} = \delta_j^N$ if $x_j=1$ and $x_i = 0$ for all $i \neq j$. An encoding function $\Encode:[0,N-1]\rightarrow \F_2^{2\ell}$ maps an integer $j \in [0, N-1]$, whose binary representation is $\itob(j) = (j_{0}, \ldots, j_{\ell-1})$, to the vector
\begin{equation*}
\Encode(j) = (1-j_0, j_0, \ldots, 1-j_i, j_i, \ldots, 1- j_{\ell-1}, j_{\ell-1}).
\end{equation*}
Given a vector $\mathbf{b} = (b_0, \ldots, b_{\ell-1}) \in \{0,1\}^\ell$, we define two pertmutations. The first permutation $T_{\mathbf{b}}: \F_2^N \rightarrow \F_2^N$ transforms $\mathbf{x} = (x_0, \ldots, x_{N-1})$ to $(x'_0, \ldots, x'_{N-1})$, where for each $i \in [0,N-1]$, we have $x_i = x'_{i^*}$ with $i^* = \btoi\big(\itob(i)\oplus \mathbf{b}\big)$. The second permutation $T'_{\mathbf{b}}: \F_2^{2\ell} \rightarrow \F_2^{2\ell}$ maps $\mathbf{f} = \left(f_0, f_1, \ldots, f_{2i}, f_{2i+1}, \ldots, f_{2(\ell-1)}, f_{2(\ell-1)+1}\right)$ to 
\[
\left(f_{b_0}, f_{1-b_0}, \ldots, f_{2i+b_i}, f_{2i+ (1-b_i)}, \ldots, f_{2(\ell-1)+ b_{\ell-1}}, f_{2(\ell-1)+ (1- b_{\ell-1})}\right).
\]
Observe that, for any $j \in [0,N-1]$ and any $\mathbf{b} \in \{0,1\}^\ell$, we have
\begin{align}
\mathbf{x} = \delta_j^N & \iff  T_{\mathbf{b}}(\mathbf{x}) = \delta^N_{\btoi(\itob(j)\oplus \mathbf{b})} \mbox{ and }\label{eqn:permutation-transformation-1} \\
\mathbf{f} = \Encode(j) & \iff  T'_{\mathbf{b}}(\mathbf{f})=\Encode(\btoi(\itob(j)\oplus \mathbf{b})).\label{eqn:permutation-transformation-2}
\end{align}

\begin{example}
Let $N=2^4$ and $j = 6$. Then $\itob(j) = (0,1,1,0)$ and $\Encode(j)= (1, 0, 0, 1, 0, 1, 1, 0)$. If $\mathbf{b} = (1, 0, 1, 0)$, then  $\btoi(\itob(j)\oplus \mathbf{b})= \btoi(1,1,0,0) = 12$. We have $T_{\mathbf{b}}(\delta_6^{16}) = \delta_{12}^{16}$ and 
\[
T'_{\mathbf{b}}(\Encode(6))= (0,1,0,1,1,0,1,0) = \Encode(12).
\]
\end{example}

\subsection{The Interactive Protocol Underlying the CPA-Anonymous Group Signature}\label{subsec:interactive-protocol}
We now present the interactive zero-knowledge argument of knowledge (\textsf{ZKAoK}) that will be used in the CPA-anonymous group signature scheme of Section~\ref{sec:main-construction}. Let $\nmc, \kmc, \tmc, m, r, \omega, \ell$ be positive integers, and $N = 2^\ell$. The public input consists of matrices $\Gmc \in \F_2^{\kmc\times\nmc}$, $\mathbf{H} \in \F_2^{r \times m}$, as well as $N$ syndromes $\y_0, \ldots, \y_{N-1}\in\F_2^r$, and vector $\mathbf{c} \in \F_2^{\nmc}$. The protocol allows the prover $\mathcal{P}$ to \emph{simultaneously} convince the verifier $\mathcal{V}$ in zero-knowledge that $\mathcal{P}$ possesses a vector
$\mathbf{s}\in \mathsf{B}(m, \omega)$ corresponding to certain syndrome $\y_j \in \{\y_0, \ldots, \y_{N-1}\}$ with hidden index $j$, \emph{and} that $\mathbf{c}$ is a correct encryption of $\itob(j)$ via the randomized McEliece encryption. Specifically, the secret witness of $\mathcal{P}$ is a tuple $(j,\mathbf{s},\mathbf{u}, \mathbf{e}) \in [0, N-1] \times \F_2^{\ncfs} \times \F_2^{\kmc - \ell} \times \F_2^{\nmc}$ that satisfies
\begin{equation}\label{eqn:relation-original}
\begin{cases}
\mathbf{H}\cdot \mathbf{s}^\top = \y_j^\top \hspace*{5pt}\wedge \hspace*{5pt} \mathbf{s} \in \mathsf{B}(m, \omega),\\
\big(\hspace*{1.5pt}\mathbf{u}\hspace*{1.5pt} \| \hspace*{1.5pt}\itob(j)\hspace*{1.5pt}\big)\cdot \Gmc \oplus \mathbf{e} = \mathbf{c}  \hspace*{5pt}\wedge \hspace*{5pt} \mathbf{e} \in \mathsf{B}(\nmc, \tmc).
\end{cases}
\end{equation}

Let $\mathbf{A} = \big[\y_0^\top | \cdots | \y_j^\top| \cdots | \y_{N\hspace*{-1.5pt}-\hspace*{-1.5pt}1}^\top \big] \in \F_2^{r \times N}$ and $\mathbf{x} = \delta_j^N$. We have $\mathbf{A}\cdot \mathbf{x}^\top = \y_j^\top$ and rewrite $\mathbf{H}\cdot \mathbf{s}^\top = \y_j^\top$ as $\mathbf{H}\cdot \mathbf{s}^\top \oplus \mathbf{A}\cdot \mathbf{x}^\top = \mathbf{0}$. Let $\widehat{\mathbf{G}} \in \F_2^{(k + \ell) \times n}$ be the matrix obtained from $\mathbf{G} \in \F_2^{k \times n}$ by replacing its last $\ell$ rows $\mathbf{g}_{k-\ell+1}, \mathbf{g}_{k -\ell + 2}, \ldots, \mathbf{g}_{k}$ by the $2\ell$ rows $\mathbf{0}^n, \mathbf{g}_{k-\ell+1}, \mathbf{0}^n, \mathbf{g}_{k -\ell + 2}, \ldots, \mathbf{0}^n, \mathbf{g}_k$. We then observe that $\big(\hspace*{1.5pt}\mathbf{u}\hspace*{1.5pt} \| \hspace*{1.5pt}\itob(j)\hspace*{1.5pt}\big)\cdot \Gmc = \big(\hspace*{1.5pt}\mathbf{u}\hspace*{1.5pt} \| \hspace*{1.5pt}\Encode(j)\hspace*{1.5pt}\big)\cdot \widehat{\Gmc}$.

Letting $\mathbf{f} = \Encode(j)$, we can equivalently rewrite (\ref{eqn:relation-original}) as
\begin{equation}\label{eqn:relation-equivalent}
\begin{cases}
\mathbf{H}\cdot \mathbf{s}^\top \oplus \mathbf{A}\cdot \mathbf{x}^\top = \mathbf{0} \hspace*{5pt}\wedge\hspace*{5pt} \mathbf{x} = \delta_j^N \hspace*{5pt}\wedge\hspace*{5pt} \mathbf{s}\in\mathsf{B}(m, \omega),\\
\big(\hspace*{1.5pt}\mathbf{u}\hspace*{1.5pt} \| \hspace*{1.5pt}\mathbf{f}\hspace*{1.5pt}\big)\cdot \widehat{\Gmc} \oplus \mathbf{e} = \mathbf{c} \hspace*{5pt}\wedge \hspace*{5pt} \mathbf{f} = \Encode(j)  \hspace*{5pt}\wedge \hspace*{5pt} \mathbf{e} \in \mathsf{B}(\nmc, \tmc).
\end{cases}
\end{equation}
To obtain a \textsf{ZKAoK} for relation (\ref{eqn:relation-equivalent}) in Stern's framework~\cite{Ste96}, $\mathcal{P}$ proceeds as follows. 

To prove that $\mathbf{x} = \delta_j^N$ \emph{and} $\mathbf{f} = \Encode(j)$ while keeping $j$ secret, $\mathcal{P}$ samples a uniformly random vector $\mathbf{b} \in \{0,1\}^\ell$, sends $\mathbf{b}_1 = \itob(j)\oplus \mathbf{b}$, and shows that
\[
T_{\mathbf{b}}(\mathbf{x}) = \delta^N_{\btoi(\mathbf{b}_1)} \hspace*{5pt}\wedge\hspace*{5pt} T'_{\mathbf{b}}(\mathbf{f})= \Encode(\btoi(\mathbf{b}_1)).
\]
By the equivalences observed in~(\ref{eqn:permutation-transformation-1}) and~(\ref{eqn:permutation-transformation-2}), the verifier will be convinced about the facts to prove. Furthermore, since $\mathbf{b}$ essentially acts as a one-time pad, the secret $j$ remains perfectly hidden.

To prove in zero-knowledge that $\mathbf{s} \in \mathsf{B}(m, \omega)$, $\mathcal{P}$ samples a uniformly random permutation $\pi \in \mathsf{S}_m$, and shows that $\pi(\mathbf{s}) \in \mathsf{B}(m, \omega)$. Similarly, to prove in zero-knowledge that $\mathbf{e} \in \mathsf{B}(\nmc, \tmc)$, a uniformly random permutation $\sigma \in \mathsf{S}_{\nmc}$ is employed.

Finally, to prove the linear equations in zero-knowledge, $\mathcal{P}$ samples uniformly random ``masking'' vectors $(\mathbf{r}_{\mathbf{s}}, \mathbf{r}_{\mathbf{x}}, \mathbf{r}_{\mathbf{u}}, \mathbf{r}_{\mathbf{f}}, \mathbf{r}_{{\mathbf{e}}})$ and shows that
\begin{align}\label{eqn:witness-masking-ch=3}
\mathbf{H}\cdot (\mathbf{s} \oplus \mathbf{r}_{\mathbf{s}})^\top \hspace*{1.5pt}\oplus \hspace*{1.5pt}\mathbf{A}\cdot (\mathbf{x} \oplus \mathbf{r}_{\mathbf{x}})^\top &= \mathbf{H}\cdot \mathbf{r}_{\mathbf{s}}^\top \hspace*{1.5pt}\oplus\hspace*{1.5pt} \mathbf{A}\cdot \mathbf{r}_{\mathbf{x}}^\top \mbox{ and } \notag\\
\big(\hspace*{1.5pt}\mathbf{u}\oplus \mathbf{r}_{\mathbf{u}}\hspace*{1.5pt} \| \hspace*{1.5pt}\mathbf{f}\oplus \mathbf{r}_{\mathbf{f}}\hspace*{1.5pt}\big)\cdot \widehat{\Gmc} \hspace*{1.5pt}\oplus \hspace*{1.5pt}({\mathbf{e}}\oplus \mathbf{r}_{{\mathbf{e}}}) \hspace*{1.5pt}\oplus \hspace*{1.5pt}\mathbf{c} &= \big(\hspace*{1.5pt}\mathbf{r}_{\mathbf{u}}\hspace*{1.5pt} \| \hspace*{1.5pt} \mathbf{r}_{\mathbf{f}}\hspace*{1.5pt}\big)\cdot \widehat{\Gmc} \hspace*{1.5pt}\oplus \hspace*{1.5pt} \mathbf{r}_{{\mathbf{e}}}.
\end{align}

Now let $\mathrm{COM}: \{0,1\}^* \rightarrow \{0,1\}^\lambda$ be a collision-resistant hash function, to be modelled as a random oracle. The prover $\mathcal{P}$ and the verifier $\mathcal{V}$ first perform the preparatory steps described above, and then interact as described in Figure~\ref{Figure:Protocol}.

\begin{figure*}[!htbp]
\begin{enumerate}
\item {\bf Commitment:} $\mathcal{P}$ samples the uniformly random objects
\begin{equation*}
\mathbf{b} \lr \{0,1\}^\ell,~ \pi \lr \mathsf{S}_m, ~ \sigma\lr \mathsf{S}_{\nmc}, ~ \rho_1, \rho_2, \rho_3 \lr \{0,1\}^\lambda, ~ 
\mathbf{r}_{\mathbf{s}}\lr\F_{2}^{\ncfs}, ~ \mathbf{r}_{\mathbf{x}} \lr \F_2^N, ~  \mathbf{r}_{\mathbf{u}}\lr\F_{2}^{\kmc-\ell}, ~ \mathbf{r}_{\mathbf{f}}\lr\F_2^{2\ell}, ~ \mathbf{r}_{{\mathbf{e}}}\lr\F_{2}^{\nmc}.
\end{equation*}
It then sends the commitment $\mathsf{CMT}:=(c_1, c_2, c_3)$ to $\mathcal{V}$, where
\begin{align*}
c_1 &= \text{COM}\big(\mathbf{b}, \hspace*{2.5pt}\pi,\hspace*{2.5pt} \sigma, \hspace*{5pt} \mathbf{H}\cdot \mathbf{r}_{\mathbf{s}}^\top \hspace*{1.5pt}\oplus \hspace*{1.5pt}\mathbf{A}\cdot \mathbf{r}_{\mathbf{x}}^\top,\hspace*{5pt} \big(\hspace*{1.5pt}\mathbf{r}_{\mathbf{u}}\hspace*{1.5pt} \| \hspace*{1.5pt} \mathbf{r}_{\mathbf{f}}\hspace*{1.5pt}\big)\cdot \widehat{\Gmc} \hspace*{1.5pt}\oplus\hspace*{1.5pt}  \mathbf{r}_{{\mathbf{e}}}; \hspace*{5.5pt} \rho_1\big),\\
c_2 &= \text{COM}\big(\pi(\mathbf{r}_{\mathbf{s}}), T_{\mathbf{b}}(\mathbf{r}_{{\mathbf{x}}}), \hspace*{2.55pt}T'_{\mathbf{b}}(\mathbf{r}_{\mathbf{f}}), \hspace*{2.55pt} \sigma(\mathbf{r}_{{\mathbf{e}}}); \hspace*{5.5pt} \rho_2\big),\\
c_3 &= \text{COM}\big(\pi(\mathbf{s}\oplus \mathbf{r}_{\mathbf{s}}), \hspace*{2.55pt} T_{\mathbf{b}}({\mathbf{x}}\oplus\mathbf{r}_{{\mathbf{x}}}), \hspace*{2.55pt}T'_{\mathbf{b}}(\mathbf{f}\oplus\mathbf{r}_{\mathbf{f}}),\hspace*{2.55pt} \sigma({\mathbf{e}}\oplus\mathbf{r}_{{\mathbf{e}}}); \hspace*{5.5pt} \rho_3\big).
\end{align*}

\item {\bf Challenge:} Upon receiving $\mathsf{CMT}$, $\mathcal{V}$ sends a challenge $\text{Ch}\lr\{1,2,3\}$ to $\mathcal{P}$.

\item {\bf Response:} $\mathcal{P}$ responds based on $\text{Ch}$.
\begin{enumerate}
\item If $\text{Ch}=1$: Reveal $c_2$ and $c_3$. Let 
\begin{equation*}
\mathbf{b}_1 = \itob(j)\oplus \mathbf{b},~
\mathbf{v}_{\mathbf{s}} = \pi(\mathbf{r}_{\mathbf{s}}),~
\mathbf{w}_{\mathbf{s}} = \pi({\mathbf{s}}),~ 
\mathbf{v}_{{\mathbf{x}}} = T_{\mathbf{b}}(\mathbf{r}_{{\mathbf{x}}}), ~ \mathbf{v}_{\mathbf{f}} = T'_{\mathbf{b}}(\mathbf{r}_{\mathbf{f}}), ~
\mathbf{v}_{{\mathbf{e}}} = \sigma(\mathbf{r}_{{\mathbf{e}}}), \mbox{ and }
\mathbf{w}_{{\mathbf{e}}} = \sigma({\mathbf{e}}).
\end{equation*}
Send $\mathsf{RSP}:=\big(\mathbf{b}_1, \hspace*{2.5pt} \mathbf{v}_{\mathbf{s}}, \hspace*{2.5pt} \mathbf{w}_{\mathbf{s}}, \hspace*{2.5pt}\mathbf{v}_{{\mathbf{x}}},
\hspace*{2.5pt}\mathbf{v}_{\mathbf{f}},
\hspace*{2.5pt}\mathbf{v}_{{\mathbf{e}}}, \hspace*{2.5pt}\mathbf{w}_{{\mathbf{e}}}; \hspace*{2.55pt} \rho_2, \rho_3 \big)$ to $\mathcal{V}$.
			
\smallskip
			
\item If $\text{Ch}=2$: Reveal $c_1$ and $c_3$. Let
\[
\mathbf{b}_2 = \mathbf{b},~ \pi_2 = \pi, ~ \sigma_2 = \sigma, ~
\mathbf{z}_{\mathbf{s}}=\mathbf{s}\oplus\mathbf{r}_{\mathbf{s}}, ~
\mathbf{z}_\mathbf{x}=\mathbf{x}\oplus\mathbf{r}_{\mathbf{x}},~
\mathbf{z}_\mathbf{u}=\mathbf{u}\oplus\mathbf{r}_{\mathbf{u}},~ \mathbf{z}_{\mathbf{f}}=\mathbf{f}\oplus\mathbf{r}_{\mathbf{f}}, \mbox{ and } \mathbf{z}_{{\mathbf{e}}} = {\mathbf{e}}\oplus\mathbf{r}_{{\mathbf{e}}}.
\]
Send $\mathsf{RSP}:=\big(\mathbf{b}_2, \hspace*{2.5pt}\pi_2, \hspace*{2.5pt}\sigma_2, \hspace*{2.5pt}\mathbf{z}_{\mathbf{s}}, \hspace*{2.5pt}\mathbf{z}_{\mathbf{x}}, \hspace*{2.5pt}\mathbf{z}_{\mathbf{u}}, \hspace*{2.5pt}\mathbf{z}_{\mathbf{f}}, \hspace*{2.5pt}\mathbf{z}_{{\mathbf{e}}}; \hspace*{2.55pt} \rho_1, \rho_3\big)$ to $\mathcal{V}$.
			
\smallskip
			
\item If $\text{Ch}=3$: Reveal $c_1$ and $c_2$. Let
\[
\mathbf{b}_3 = \mathbf{b},~ \pi_3 = \pi,~ \sigma_3 = \sigma,~ 
\mathbf{y}_{\mathbf{s}}=\mathbf{r}_{\mathbf{s}},~
\mathbf{y}_\mathbf{x}=\mathbf{r}_{\mathbf{x}},~
\mathbf{y}_\mathbf{u}=\mathbf{r}_{\mathbf{u}},~
\mathbf{y}_{\mathbf{f}}=\mathbf{r}_{\mathbf{f}}, \mbox{ and }
\mathbf{y}_{{\mathbf{e}}} = \mathbf{r}_{{\mathbf{e}}}.
\]
Send $\mathsf{RSP}:=\big(\mathbf{b}_3, \hspace*{2.5pt}\pi_3, \hspace*{2.5pt}\sigma_3, \hspace*{2.5pt}\mathbf{y}_{\mathbf{s}}, \hspace*{2.5pt}\mathbf{y}_{\mathbf{x}}, \hspace*{2.5pt}\mathbf{y}_{\mathbf{u}}, \hspace*{2.5pt}\mathbf{y}_{\mathbf{f}}, \hspace*{2.5pt}\mathbf{y}_{{\mathbf{e}}}; \hspace*{2.55pt} \rho_1, \rho_2\big)$ to $\mathcal{V}$.
\end{enumerate}

\item {\bf Verification:} Upon receiving $\mathsf{RSP}$, $\mathcal{V}$ proceeds based on $\text{Ch}$.
\begin{enumerate}
\item If $\text{Ch}=1$: Let $\mathbf{w}_{\mathbf{x}} = \delta^N_{\btoi(\mathbf{b}_1)} \in \F_2^N$ and   $\mathbf{w}_{\mathbf{f}}=\Encode(\btoi(\mathbf{b}_1)) \in \F_2^{2\ell}$.
		
Check that ${\mathbf{w}}_{\mathbf{s}}\in \mathsf{B}(m, \omega)$, 
$\mathbf{w}_{{\mathbf{e}}} \in \mathsf{B}(\nmc, \tmc)$, 
\[
c_2 = \text{COM}\big({\mathbf{v}}_{\mathbf{s}}, \hspace*{5pt}{\mathbf{v}}_{\mathbf{x}}, \hspace*{5pt}{\mathbf{v}}_{\mathbf{f}}, \hspace*{5pt}{\mathbf{v}}_{{\mathbf{e}}}; \hspace*{5pt} \rho_2\big) \mbox{, and }
c_3 = \text{COM}\big({\mathbf{v}}_{\mathbf{s}}\oplus{\mathbf{w}}_{\mathbf{s}}, \hspace*{5pt}{\mathbf{v}}_{\mathbf{x}}\oplus{\mathbf{w}}_{\mathbf{x}}, \hspace*{5pt}{\mathbf{v}}_{\mathbf{f}}\oplus{\mathbf{w}}_{\mathbf{f}},\hspace*{5pt} {\mathbf{v}}_{{\mathbf{e}}}\oplus{\mathbf{w}}_{{\mathbf{e}}}; \hspace*{5pt} \rho_3\big).
\]

\item If $\text{Ch}=2$: Check that
\begin{align*}
c_1 &= \text{COM}\big(\mathbf{b}_2,\hspace*{5pt} \pi_2, \hspace*{5pt}\sigma_2, \hspace*{5pt}\Hcfs\cdot\mathbf{z}_{\mathbf{s}}^\top \hspace*{1.5pt}\oplus \hspace*{1.5pt}\mathbf{A}\cdot \mathbf{z}_{\mathbf{x}}^\top, \hspace*{5pt}\big(\hspace*{1.5pt}\mathbf{z}_{\mathbf{u}}\hspace*{1.5pt}\|\hspace*{1.5pt}\mathbf{z}_{\mathbf{f}}\hspace*{1.5pt}\big)\cdot\widehat{\Gmc}\hspace*{1.5pt}\oplus\hspace*{1.5pt}\mathbf{z}_{{\mathbf{e}}} \hspace*{1.5pt}\oplus\hspace*{1.5pt} \mathbf{c}; \hspace*{5pt} \rho_1\big) \mbox{ and }\\
c_3 &= \text{COM}\big(\pi_2(\mathbf{z}_{\mathbf{s}}), \hspace*{5pt}T_{\mathbf{b}_2}(\mathbf{z}_{{\mathbf{x}}}), \hspace*{5pt}T'_{\mathbf{b}_2}(\mathbf{z}_{\mathbf{f}}\big), \hspace*{5pt} \sigma_2(\mathbf{z}_{{\mathbf{e}}}); \hspace*{5pt} \rho_3\big).
\end{align*}

\item If $\text{Ch}=3$: Check that
\begin{align*}
c_1 &= \text{COM}\big(\mathbf{b}_3,\hspace*{5pt} \pi_3, \hspace*{5pt}\sigma_3, \hspace*{5pt}\Hcfs\cdot\mathbf{y}_{\mathbf{s}}^\top \hspace*{1.5pt}\oplus \hspace*{1.5pt}\mathbf{A}\cdot \mathbf{y}_{\mathbf{x}}^\top , \hspace*{5pt}\big(\hspace*{1.5pt}\mathbf{y}_{\mathbf{u}}\hspace*{1.5pt}\|\hspace*{1.5pt}\mathbf{y}_{\mathbf{f}}\hspace*{1.5pt}\big)\cdot\widehat{\Gmc}\hspace*{1.5pt}\oplus\hspace*{1.5pt}\mathbf{y}_{{\mathbf{e}}}; \hspace*{5pt} \rho_1\big) \mbox{ and }\\
c_2 &= \text{COM}\big(\pi_3(\mathbf{y}_{\mathbf{s}}), \hspace*{5pt} T_{\mathbf{b}_3}(\mathbf{y}_{{\mathbf{x}}}), \hspace*{5pt}T'_{\mathbf{b}_3}(\mathbf{y}_{\mathbf{f}}), \hspace*{5pt} \sigma_3(\mathbf{y}_{{\mathbf{e}}}); \hspace*{5pt} \rho_2\big).
\end{align*}
\end{enumerate}
In each case, $\mathcal{V}$ outputs $1$ if and only if all of the conditions hold. Otherwise, $\mathcal{V}$ outputs $0$.
\end{enumerate}

\caption{The underlying ZK protocol of the CPA-anonymous group signature.}
\label{Figure:Protocol}
\end{figure*}

\smallskip

\noindent
{\bf Analysis of the Protocol. }
The following theorem summarizes the properties of our protocol.

\begin{theorem}\label{theorem:Protocol-properties}
The interactive protocol described in Section~\ref{subsec:interactive-protocol} has perfect completeness. Its communication cost is bounded above by $\beta= (N + 3\log N) + m(\log m +1) + n(\log n +1) + k + 5\lambda$ bits. If $\mathrm{COM}$ is modelled as a random oracle, then the protocol is statistical zero-knowledge. If $\mathrm{COM}$ is a collision-resistant hash function, then the protocol is an argument of knowledge.
\end{theorem}

The given interactive protocol is perfectly {\bf complete}, \ie, if $\mathcal{P}$ possesses a valid witness $(j , \mathbf{s}, \mathbf{u}, \mathbf{e})$ and follows the protocol, then $\mathcal{V}$ always outputs $1$. Indeed, given $(j , \mathbf{s}, \mathbf{u}, \mathbf{e})$ satisfying~(\ref{eqn:relation-original}), $\mathcal{P}$ can always obtain $(j, \mathbf{s}, \mathbf{x}, \mathbf{u}, \mathbf{f}, {\mathbf{e}})$ satisfying~(\ref{eqn:relation-equivalent}). Then, as discussed above, the following three assertions hold.
\begin{align*}
& \forall ~ \pi \in \mathsf{S}_m: ~\pi(\mathbf{s})\in \mathsf{B}(m, \omega), \\ 
&\forall ~ \sigma \in \mathsf{S}_{\nmc}: ~ \sigma({\mathbf{e}})\in \mathsf{B}(\nmc, \tmc), \\
& \forall ~ \mathbf{b} \in \{0,1\}^\ell: T_{\mathbf{b}}(\mathbf{x}) = \delta^N_{\btoi(\itob(j)\oplus \mathbf{b})} = \mathbf{w_x} \mbox{ and } T'_{\mathbf{b}}(\mathbf{f}) = \Encode(\btoi(\itob(j)\hspace*{-1.25pt} \oplus \mathbf{b})) = \mathbf{w_f}.
\end{align*}
Thus, $\mathcal{P}$ always passes $\mathcal{V}$'s checks whenever $\text{Ch}=1$. When $\text{Ch}=2$, $\mathcal{P}$ also passes the verification since the linear equations in~(\ref{eqn:witness-masking-ch=3}) hold true. Finally, for $\text{Ch}=3$, it suffices to note that $\mathcal{V}$ simply checks for honest computations of $c_1$ and $c_2$.

Let us now consider the {\bf communication cost}. The commitment \textsf{CMT} has bit-size $3\lambda$. If $\mathrm{Ch}=1$, then the response \textsf{RSP} has bit-size $3\ell + N + 2(m +n + \lambda)$. If $\mathrm{Ch}=2$ or $\mathrm{Ch}=3$, then \textsf{RSP} has bit-size $2\ell + N + m(\log m + 1) + n(\log n + 1) + k + 2\lambda$. Therefore, the protocol's total communication cost, in bits, is less than the specified bound $\beta$. 

The following lemma says that our interactive protocol is {\bf statistically zero-knowledge} if COM is modelled as a random oracle. 
It employs the standard simulation technique for Stern-type protocols as was done, \eg, in~\cite{Ste96,KTX08}, and \cite{LNSW13}.
\begin{lemma}\label{lemma:proof-of-zero-knowledge}
In the random oracle model, there exists an efficient simulator $\mathcal{S}$ interacting with a (possibly cheating) verifier $\widehat{\mathcal{V}}$, such that, given only the public input of the protocol, $\mathcal{S}$ outputs, with probability negligibly close to $2/3$, a simulated transcript that is statistically close to the one produced by the honest prover in the real interaction.
\end{lemma}
\begin{IEEEproof}
Simulator $\mathcal{S}$, given the public input $(\mathbf{H}, \mathbf{A}, \widehat{\mathbf{G}}, \mathbf{c})$, begins by selecting a random $\overline{\text{Ch}} \in \{1,2,3\}$. This is a prediction of the challenge value that$\widehat{\mathcal{V}}$ will \emph{not} choose.

\begin{enumerate}
\item \textbf{Case }$\overline{\text{Ch}}=1$: $\mathcal{S}$ proceeds as follows.
\begin{enumerate}
\item Compute $\mathbf{s}' \in \F_2^{\ncfs}$ and $\mathbf{x}' \in \F_2^N$ such that $\mathbf{H}\cdot \mathbf{s}'^\top\oplus \mathbf{A}\cdot\mathbf{x}'^\top = \mathbf{0}$. Compute $\mathbf{u}' \in \F_2^{\kmc -\ell}$, $\mathbf{f}' \in \F_2^{2\ell}$, ${\mathbf{e}}' \in \F_2^{\nmc}$ such that $\big(\hspace*{1.5pt}\mathbf{u}' \hspace*{1.5pt}\| \hspace*{1.5pt}\mathbf{f}'\hspace*{1.5pt}\big)\cdot \widehat{\mathbf{G}} \oplus {\mathbf{e}}'= \mathbf{c}$. These steps can be done efficiently by using linear algebraic tools.
\item Sample uniformly random objects, and send a commitment computed in the same manner as of the real prover. More explicitly, $\mathcal{S}$ samples
\begin{align*}
&\mathbf{b} \lr \{0,1\}^\ell, \quad 
\pi \lr \mathsf{S}_m, \quad
\sigma\lr \mathsf{S}_{\nmc}, \quad 
\rho_1, \rho_2, \rho_3 \lr \{0,1\}^\lambda, \\
&\mathbf{r}_{\mathbf{s}}\lr\F_{2}^{\ncfs}, \quad 
\mathbf{r}_{\mathbf{x}} \lr \F_2^N, \quad \mathbf{r}_{\mathbf{u}}\lr\F_{2}^{\kmc-\ell}, \quad \mathbf{r}_{\mathbf{f}}\lr\F_2^{2\ell}, \quad  \mathbf{r}_{{\mathbf{e}}}\lr\F_{2}^{\nmc},
\end{align*}
and sends the commitment $\mathsf{CMT}:=(c_1^{'}, c_2^{'}, c_3^{'})$, where
\begin{align}\label{eqn:cmt-simulation}
c_1^{'} &= \text{COM}\big(\mathbf{b}, \hspace*{2.5pt}\pi,\hspace*{2.5pt} \sigma, \hspace*{5pt} \mathbf{H}\cdot \mathbf{r}_{\mathbf{s}}^\top \hspace*{1.5pt}\oplus \hspace*{1.5pt}\mathbf{A}\cdot \mathbf{r}_{\mathbf{x}}^\top,\hspace*{5pt} \big(\hspace*{1.5pt}\mathbf{r}_{\mathbf{u}}\hspace*{1.5pt} \| \hspace*{1.5pt} \mathbf{r}_{\mathbf{f}}\hspace*{1.5pt}\big)\cdot \widehat{\Gmc} \hspace*{1.5pt}\oplus\hspace*{1.5pt}  \mathbf{r}_{{\mathbf{e}}}; \hspace*{5.5pt} \rho_1\big),\notag \\
c_2^{'} &= \text{COM}\big(\pi(\mathbf{r}_{\mathbf{s}}), T_{\mathbf{b}}(\mathbf{r}_{{\mathbf{x}}}), \hspace*{2.55pt}T'_{\mathbf{b}}(\mathbf{r}_{\mathbf{f}}), \hspace*{2.55pt} \sigma(\mathbf{r}_{{\mathbf{e}}}); \hspace*{5.5pt} \rho_2\big),\notag \\
c_3^{'} &= \text{COM}\big(\pi(\mathbf{s}'\oplus \mathbf{r}_{\mathbf{s}}), \hspace*{2.55pt} T_{\mathbf{b}}({\mathbf{x}'}\oplus\mathbf{r}_{{\mathbf{x}}}), \hspace*{2.55pt}T'_{\mathbf{b}}(\mathbf{f}'\oplus\mathbf{r}_{\mathbf{f}}),\hspace*{2.55pt} \sigma({\mathbf{e}'}\oplus\mathbf{r}_{{\mathbf{e}}}); \hspace*{5.5pt} \rho_3\big).
\end{align}
\end{enumerate}
Upon receiving a challenge $\text{Ch}$ from $\widehat{\mathcal{V}}$, the simulator responds accordingly.
\begin{enumerate}
	\item If $\text{Ch}=1$: Output $\perp$ and abort.
	\item If $\text{Ch}=2$: Send $\mathsf{RSP}=\big(\mathbf{b}, \hspace*{2.5pt}\pi, \hspace*{2.5pt}\sigma, \hspace*{2.5pt}\mathbf{s}'\oplus\mathbf{r}_{\mathbf{s}}, \hspace*{2.5pt} \mathbf{x}' \oplus \mathbf{r}_{\mathbf{x}},  \hspace*{2.5pt}\mathbf{u}'\oplus\mathbf{r}_{\mathbf{u}}, \hspace*{2.5pt}\mathbf{f}'\oplus\mathbf{r}_{\mathbf{f}}, \hspace*{2.5pt}{\mathbf{e}}'\oplus\mathbf{r}_{{\mathbf{e}}}; \hspace*{5pt} \rho_1, \rho_3\big)$.
	\item If $\text{Ch}=3$: Send $\mathsf{RSP}=\big(\mathbf{b}, \hspace*{2.5pt}\pi, \hspace*{2.5pt}\sigma, \hspace*{2.5pt}\mathbf{r}_{\mathbf{s}}, \hspace*{2.5pt} \mathbf{r}_{\mathbf{x}}, \hspace*{2.5pt}\mathbf{r}_{\mathbf{u}}, \hspace*{2.5pt}\mathbf{r}_{\mathbf{f}}, \hspace*{2.5pt}\mathbf{r}_{{\mathbf{e}}}; \hspace*{5pt} \rho_1, \rho_2\big)$.
\end{enumerate}

\item \textbf{Case }$\overline{\text{Ch}}=2$: $\mathcal{S}$ samples
\begin{align*}
& j' \lr [0, N-1], \quad 
\mathbf{s}' \lr \mathsf{B}(m,\omega), \quad 
{\mathbf{e}}' \lr \mathsf{B}(\nmc, \tmc), \quad 
\mathbf{b} \lr \{0,1\}^\ell, \quad 
\pi \lr \mathsf{S}_m, \quad
\sigma\lr \mathsf{S}_{\nmc},\\
&\rho_1, \rho_2, \rho_3 \lr \{0,1\}^\lambda, \quad 
\mathbf{r}_{\mathbf{s}}\lr\F_{2}^{\ncfs}, \quad
\mathbf{r}_{\mathbf{x}} \lr \F_2^N, \quad \mathbf{r}_{\mathbf{u}}\lr\F_{2}^{\kmc-\ell}, \quad
\mathbf{r}_{\mathbf{f}}\lr\F_2^{2\ell}, \quad  \mathbf{r}_{{\mathbf{e}}}\lr\F_{2}^{\nmc}.
\end{align*}
It also lets $\mathbf{x}' = \delta^N_{j'}$ and $\mathbf{f}' = \Encode(j')$.
Then $\mathcal{S}$ sends the commitment $\mathsf{CMT}$ computed in the same manner as in~(\ref{eqn:cmt-simulation}).

\smallskip
	
\noindent
Upon receiving a challenge $\text{Ch}$ from $\widehat{\mathcal{V}}$, it responds as follows.
\begin{enumerate}
\item If $\text{Ch}=1$: Send $\mathsf{RSP}=\big(\itob(j')\oplus \mathbf{b}, \hspace*{1.5pt} \pi(\mathbf{r}_{\mathbf{s}}), \hspace*{1.5pt} \pi(\mathbf{s}'),  \hspace*{1.5pt}T_{\mathbf{b}}(\mathbf{r}_{{\mathbf{x}}}), \hspace*{1.5pt}T'_{\mathbf{b}}(\mathbf{r}_{\mathbf{f}}), \hspace*{1.5pt}\sigma(\mathbf{r}_{{\mathbf{e}}}), \hspace*{1.5pt}\sigma({\mathbf{e}}'); \hspace*{2.5pt} \rho_2, \rho_3\big)$.
\item If $\text{Ch}=2$: Output $\perp$ and abort.
\item If $\text{Ch}=3$: Send $\mathsf{RSP}$ computed as in the case $(\overline{\text{Ch}}=1, \text{Ch}=3)$.
\end{enumerate}
	
\item \textbf{Case }$\overline{\text{Ch}}=3$: The simulator performs the preparation as in the case $\overline{\text{Ch}}=2$ above. Additionally, it samples $\mathbf{u}' \lr \F_2^{\kmc - \ell}$.
It then sends the commitment $\mathsf{CMT}:= (c_1^{'}, c_2^{'}, c_3^{'})$, where $c_2^{'}, c_3^{'}$ are computed as in~(\ref{eqn:cmt-simulation}) and 
\[
c_1^{'} \hspace*{-1.5pt}= \hspace*{-1.5pt}\text{COM}\big(\mathbf{b}, \pi, \sigma, \hspace*{1pt} {\Hcfs}\cdot(\mathbf{s}'\oplus\mathbf{r}_{\mathbf{s}})^\top \oplus \mathbf{A}\cdot(\mathbf{x}' \oplus \mathbf{r}_{\mathbf{x}})^\top\hspace*{-2.5pt},\big(\mathbf{u}' \oplus \mathbf{r}_{\mathbf{u}}\|\mathbf{f}' \oplus \mathbf{r}_{\mathbf{f}}\big)\cdot\widehat{\Gmc}\oplus\big({\mathbf{e}}'\oplus \mathbf{r}_{{\mathbf{e}}}\big) \oplus \mathbf{c}; \hspace*{0.5pt} \rho_1\big).
\]
Upon receiving a challenge $\text{Ch}$ from $\widehat{\mathcal{V}}$, it responds as follows.
\begin{enumerate}
\item If $\text{Ch}=1$: Send $\mathsf{RSP}$ computed as in the case $(\overline{\text{Ch}}=2, \text{Ch}=1)$.
\item If $\text{Ch}=2$: Send $\mathsf{RSP}$ computed as in the case $(\overline{\text{Ch}}=1, \text{Ch}=2)$.
\item If $\text{Ch}=3$: Output $\perp$ and abort.
\end{enumerate}
\end{enumerate}
	
In every case that we have considered above, the distribution of the commitment $\mathsf{CMT}$ and the distribution of the challenge $\text{Ch}$ from $\widehat{\mathcal{V}}$ are statistically close to those in the real interaction, since the outputs of the random oracle COM are assumed to be uniformly random. Hence, the probability that $\mathcal{S}$ outputs $\perp$ is negligibly close to $1/3$. Moreover, one can check that whenever the simulator does not abort, it provides a successful transcript, whose distribution is statistically close to that of the prover in the real interaction. We have thus constructed a simulator that can successfully impersonate the honest prover, with probability $2/3$.  

\end{IEEEproof}

The next lemma 
establishes that our protocol satisfies the special soundness property of $\Sigma$-protocols. This implies, as had been shown in ~\cite{Groth04}, that the protocol is indeed an {\bf argument of knowledge}.
\begin{lemma}\label{lemma:proof-of-knowledge-extraction}
Let $\mathrm{COM}$ be a collision-resistant hash function. Given the public input of the protocol, a commitment $\mathsf{CMT}$ and three valid responses $\mathsf{RSP}_1, \mathsf{RSP}_2, \mathsf{RSP}_3$ to all three possible values of the challenge $\mathrm{Ch}$, one can efficiently construct a knowledge extractor $\mathcal{E}$ that outputs a tuple $(j', \mathbf{s}', \mathbf{u}', \mathbf{e}')\in [0, N-1] \times \F_2^{\ncfs} \times \F_2^{\kmc - \ell} \times \F_2^{\nmc}$ that simultaneously satisfies the requirements
$\mathbf{H}\cdot \mathbf{s}'^\top = \y_{j'}^\top, ~ 
\mathbf{s}'\in\mathsf{B}(m,\omega), ~
\big(\hspace*{1.5pt}\mathbf{u}'\hspace*{1.5pt} \| \hspace*{1.5pt}\itob(j')\hspace*{1.5pt}\big)\cdot \Gmc \hspace*{1.5pt}\oplus \hspace*{1.5pt}\mathbf{e}' = \mathbf{c} \mbox{, and } 
\mathbf{e}'\in \mathsf{B}(\nmc, \tmc)$.
\end{lemma}
\begin{IEEEproof}
Assume that we have a commitment $\mathsf{CMT} = (c_1, c_2, c_3)$ and the three responses
\begin{align*}
\mathsf{RSP}_1 &=\big(\mathbf{b}_1, \hspace*{2.5pt} \mathbf{v}_{\mathbf{s}}, \hspace*{2.5pt} \mathbf{w}_{\mathbf{s}}, \hspace*{2.5pt}\mathbf{v}_{{\mathbf{x}}}, \hspace*{2.5pt}\mathbf{v}_{\mathbf{f}}, \hspace*{2.5pt}\mathbf{v}_{{\mathbf{e}}}, \hspace*{2.5pt}\mathbf{w}_{{\mathbf{e}}}; \hspace*{2.55pt} \rho_2, \rho_3 \big),\\
\mathsf{RSP}_2 &=\big(\mathbf{b}_2, \hspace*{2.5pt}\pi_2, \hspace*{2.5pt}\sigma_2, \hspace*{2.5pt}\mathbf{z}_{\mathbf{s}}, \hspace*{2.5pt}\mathbf{z}_{\mathbf{x}}, \hspace*{2.5pt}\mathbf{z}_{\mathbf{u}}, \hspace*{2.5pt}\mathbf{z}_{\mathbf{f}}, \hspace*{2.5pt}\mathbf{z}_{{\mathbf{e}}}; \hspace*{2.55pt} \rho_1, \rho_3\big),\\
\mathsf{RSP}_3 &= \big(\mathbf{b}_3, \hspace*{2.5pt}\pi_3, \hspace*{2.5pt}\sigma_3, \hspace*{2.5pt}\mathbf{y}_{\mathbf{s}}, \hspace*{2.5pt}\mathbf{y}_{\mathbf{x}}, \hspace*{2.5pt}\mathbf{y}_{\mathbf{u}}, \hspace*{2.5pt}\mathbf{y}_{\mathbf{f}}, \hspace*{2.5pt}\mathbf{y}_{{\mathbf{e}}}; \hspace*{2.55pt} \rho_1, \rho_2\big)
\end{align*}
that satisfy all of the verification conditions when $\text{Ch}=1$, $\text{Ch}=2$, and $\text{Ch}=3$, respectively. More explicitly, we have the relations
\begin{align*}
{\mathbf{w}}_{\mathbf{s}} &\in \mathsf{B}(\ncfs, \tcfs), \hspace*{2.5pt} \mathbf{w}_{\mathbf{x}} = \delta^N_{\btoi(\mathbf{b}_1)},  \hspace*{2.5pt} \mathbf{w}_{\mathbf{f}}=\Encode(\btoi(\mathbf{b}_1)), \hspace*{2.5pt} \mathbf{w}_{{\mathbf{e}}} \in \mathsf{B}(\nmc, \tmc),\\
c_1 &= \text{COM}\big(\mathbf{b}_2,\hspace*{2.5pt} \pi_2, \hspace*{2.5pt}\sigma_2, \hspace*{2.5pt}\Hcfs\cdot\mathbf{z}_{\mathbf{s}}^\top \hspace*{1.5pt}\oplus \hspace*{1.5pt}\mathbf{A}\cdot \mathbf{z}_{\mathbf{x}}^\top, \hspace*{2.5pt}\big(\hspace*{1.5pt}\mathbf{z}_{\mathbf{u}}\hspace*{1.5pt}\|\hspace*{1.5pt}\mathbf{z}_{\mathbf{f}}\hspace*{1.5pt}\big)\cdot\widehat{\Gmc}\hspace*{1.5pt}\oplus\hspace*{1.5pt}\mathbf{z}_{{\mathbf{e}}}  \hspace*{1.5pt}\oplus\hspace*{1.5pt} \mathbf{c}; \hspace*{2.5pt} \rho_1\big)\\
&= \text{COM}\big(\mathbf{b}_3,\hspace*{2.5pt} \pi_3, \hspace*{2.5pt}\sigma_3, \hspace*{2.5pt}\Hcfs\cdot\mathbf{y}_{\mathbf{s}}^\top \hspace*{1.5pt}\oplus \hspace*{1.5pt}\mathbf{A}\cdot \mathbf{y}_{\mathbf{x}}^\top , \hspace*{2.5pt}\big(\hspace*{1.5pt}\mathbf{y}_{\mathbf{u}}\hspace*{1.5pt}\|\hspace*{1.5pt}\mathbf{y}_{\mathbf{f}}\hspace*{1.5pt}\big)\cdot\widehat{\Gmc}\hspace*{1.5pt}\oplus\hspace*{1.5pt}\mathbf{y}_{{\mathbf{e}}}; \hspace*{2.5pt} \rho_1\big),\\
c_2 &= \text{COM}\big({\mathbf{v}}_{\mathbf{s}}, \hspace*{1pt}{\mathbf{v}}_{\mathbf{x}}, \hspace*{1pt}{\mathbf{v}}_{\mathbf{f}}, \hspace*{1pt}{\mathbf{v}}_{{\mathbf{e}}}; \hspace*{1.5pt} \rho_2\big) \hspace*{-1.5pt}=\hspace*{-1.5pt} \text{COM}\big(\pi_3(\mathbf{y}_{\mathbf{s}}), \hspace*{1pt} T_{\mathbf{b}_3}(\mathbf{y}_{{\mathbf{x}}}), \hspace*{1pt}T'_{\mathbf{b}_3}(\mathbf{y}_{\mathbf{f}}), \hspace*{1pt} \sigma_3(\mathbf{y}_{{\mathbf{e}}}); \hspace*{1.5pt} \rho_2\big),\\
c_3 &= \text{COM}\big({\mathbf{v}}_{\mathbf{s}}\oplus{\mathbf{w}}_{\mathbf{s}}, \hspace*{2.5pt}{\mathbf{v}}_{\mathbf{x}}\oplus{\mathbf{w}}_{\mathbf{x}}, \hspace*{2.5pt}{\mathbf{v}}_{\mathbf{f}}\oplus{\mathbf{w}}_{\mathbf{f}},\hspace*{2.5pt} {\mathbf{v}}_{{\mathbf{e}}}\oplus{\mathbf{w}}_{{\mathbf{e}}}; \hspace*{5pt} \rho_3\big)
= \text{COM}\big(\pi_2(\mathbf{z}_{\mathbf{s}}), \hspace*{2.5pt}T_{\mathbf{b}_2}(\mathbf{z}_{{\mathbf{x}}}), \hspace*{2.5pt}T'_{\mathbf{b}_2}(\mathbf{z}_{\mathbf{f}}\big), \hspace*{2.5pt} \sigma_2(\mathbf{z}_{{\mathbf{e}}}); \hspace*{5pt} \rho_3\big).
\end{align*}
Based on the collision-resistance property of COM, we can infer that
\begin{align*}
&\mathbf{b}_2 \hspace*{-1pt}=\hspace*{-1pt} \mathbf{b}_3; \hspace*{0.5pt}\pi_2 = \pi_3, \quad \sigma_2 \hspace*{-1pt}= \hspace*{-1pt} \sigma_3;\hspace*{0.5pt} \delta^N_{\btoi(\mathbf{b}_1)} \hspace*{-1.5pt}= \hspace*{-1pt}{\mathbf{w}}_{{\mathbf{x}}} = T_{\mathbf{b}_2}(\mathbf{z}_{{\mathbf{x}}}) \hspace*{-1pt}\oplus\hspace*{-1pt} T_{ \mathbf{b}_3}(\mathbf{y}_{{\mathbf{x}}}) \hspace*{-1pt}= \hspace*{-1pt} T_{\mathbf{b}_2}(\mathbf{z}_{{\mathbf{x}}} \hspace*{-1pt}\oplus\hspace*{-1pt} \mathbf{y}_{{\mathbf{x}}}), \\
&\Encode(\btoi(\mathbf{b}_1))= \mathbf{w}_{\mathbf{f}}= T'_{\mathbf{b}_2}(\mathbf{z}_{\mathbf{f}}) \oplus T'_{\mathbf{b}_3}(\mathbf{y}_{\mathbf{f}}) =  T'_{\mathbf{b}_2}(\mathbf{z}_{\mathbf{f}} \oplus \mathbf{y}_{\mathbf{f}}),\\
&\mathsf{B}(\ncfs, \tcfs) \ni {\mathbf{w}}_{\mathbf{s}} =  \pi_2(\mathbf{z}_{\mathbf{s}})\oplus \pi_3(\mathbf{y}_{\mathbf{s}}) = \pi_2(\mathbf{z}_{\mathbf{s}} \oplus \mathbf{y}_{\mathbf{s}}),\\
&\mathsf{B}(\nmc, \tmc) \ni \mathbf{w}_{{\mathbf{e}}} = \sigma_2(\mathbf{z}_{{\mathbf{e}}}) \oplus \sigma_3(\mathbf{y}_{{\mathbf{e}}}) = \sigma_2(\mathbf{z}_{{\mathbf{e}}} \oplus \mathbf{y}_{{\mathbf{e}}}),\\
&\Hcfs\hspace*{-1.5pt}\cdot\hspace*{-1.5pt}(\mathbf{z}_{\mathbf{s}} \oplus \mathbf{y}_{\mathbf{s}})^\top \oplus \mathbf{A}\hspace*{-1.5pt}\cdot\hspace*{-1.5pt}(\mathbf{z}_{{\mathbf{x}}} \oplus \mathbf{y}_{{\mathbf{x}}})^\top = \mathbf{0} \mbox{, and }  \big(\mathbf{z}_{\mathbf{u}} \oplus \mathbf{y}_{\mathbf{u}} \|\mathbf{z}_{\mathbf{f}} \oplus \mathbf{y}_{\mathbf{f}}\big)\hspace*{-1.5pt}\cdot\hspace*{-1.5pt} \widehat{\mathbf{G}} \oplus (\mathbf{z}_{{\mathbf{e}}} \oplus \mathbf{y}_{{\mathbf{e}}}) = \mathbf{c}.
\end{align*}

Let $j' = \btoi(\mathbf{b}_1 \oplus \mathbf{b}_2) \in [0, N-1]$. Let ${\mathbf{x}}' = \mathbf{z}_{{\mathbf{x}}} \oplus \mathbf{y}_{{\mathbf{x}}} \in \F_2^{N}$. Then, by~(\ref{eqn:permutation-transformation-1}), we have 
${\mathbf{x}}' =  \delta_{j'}^N$. Thus, $\mathbf{A}\cdot\mathbf{x}'^\top = \mathbf{y}_{j'}^\top$. 

Let $\mathbf{f}' = \mathbf{z}_{\mathbf{f}} \oplus \mathbf{y}_{\mathbf{f}} \in \F_2^\ell$. Then, by~(\ref{eqn:permutation-transformation-2}), 
we have $\mathbf{f}' = \Encode(j')$. 

Let $\mathbf{s}' = \mathbf{z}_{\mathbf{s}} \oplus \mathbf{y}_{\mathbf{s}} \in \F_2^{\ncfs}$. Then we have $\mathbf{s}' = \pi_2^{-1}({\mathbf{w}}_{\mathbf{s}}) \in \mathsf{B}(\ncfs,\tcfs)$. 

Let ${\mathbf{e}}' = \mathbf{z}_{{\mathbf{e}}} \oplus \mathbf{y}_{{\mathbf{e}}} \in \F_2^{\nmc}$. Then we have ${\mathbf{e}}' = \sigma_2^{-1}({\mathbf{w}}_{{\mathbf{e}}}) \in \mathsf{B}(\nmc, \tmc)$. Let $\mathbf{u}' = \mathbf{z}_{\mathbf{u}} \oplus \mathbf{y}_{\mathbf{u}} \in \F_2^{\kmc - \ell}$. 

Furthermore, we have 
$\Hcfs\cdot\mathbf{s}'^\top \oplus \mathbf{A}\cdot\mathbf{x}'^\top = \mathbf{0}$ and $\big(\mathbf{u}' \|\Encode(j')\big)\cdot \widehat{\mathbf{G}} \oplus {\mathbf{e}}' = \mathbf{c}$. They imply, respectively, that ${\Hcfs}\cdot{\mathbf{s}}'^\top = \mathbf{A}\cdot\mathbf{x}'^\top = \y_{j'}^\top$ and $\big(\mathbf{u}' \|\itob(j')\big)\cdot \mathbf{G} \oplus {\mathbf{e}}' = \mathbf{c}$.
	
We have thus constructed an efficient extractor $\mathcal{E}$ that outputs 
$(j', \mathbf{s}', \mathbf{u}', \mathbf{e}')\in [0, N-1] \times \F_2^{\ncfs} \times \F_2^{\kmc - \ell} \times \F_2^{\nmc}$ satisfying
\[
\mathbf{H}\cdot \mathbf{s}'^\top = \y_{j'}^\top, \quad 
\mathbf{s}'\in\mathsf{B}(m,\omega), \quad
\big(\hspace*{1.5pt}\mathbf{u}'\hspace*{1.5pt} \| \hspace*{1.5pt}\itob(j')\hspace*{1.5pt}\big)\cdot \Gmc \oplus \mathbf{e}' = \mathbf{c}, \quad 
\mathbf{e}'\in \mathsf{B}(\nmc, \tmc).
\]
This completes the proof. 
\end{IEEEproof}

\section{The Interactive Protocol Underlying the CCA-Anonymous Group Signature}\label{sec:appendix-CCA-ZK-protocol}

The required ZK protocol is a simple extension of the one underlying the CPA-anonymous group signature that we have described in Section~\ref{sec:nizk}.  Here, we handle two ciphertexts $\mathbf{c}^{(1)}$ and $\mathbf{c}^{(2)}$ of $\itob(j)$ by executing two instances of the techniques used for handling one ciphertext $\mathbf{c}$ from Section~\ref{sec:nizk}.  

Applying the same transformations as in Section~\ref{sec:nizk}, we can translate the statement to be proved to proving knowledge of $\mathbf{s}, \mathbf{x}, \{\mathbf{u}^{(i)}\}_{i \in [2]}, \mathbf{f}, \{\mathbf{e}^{(i)}\}_{i \in [2]}$ such that 
\begin{align*}
&\mathbf{H}\cdot \mathbf{s}^\top \oplus \mathbf{A}\cdot \mathbf{x}^\top = \mathbf{0}, \quad
\mathbf{x} = \delta_j^N, \quad 
\mathbf{s}\in\mathsf{B}(m, \omega),\\
&\{\big(\hspace*{1.5pt}\mathbf{u}^{(i)}\hspace*{1.5pt} \| \hspace*{1.5pt}\mathbf{f}\hspace*{1.5pt}\big)\cdot \widehat{\Gmc}^{(i)} \oplus \mathbf{e}^{(i)} = \mathbf{c}^{(i)}\}_{i \in [2]}, \quad 
\mathbf{f} = \Encode(j), \quad
\{\mathbf{e}^{(i)} \in \mathsf{B}(\nmc, \tmc)\}_{i \in [2]}, 
\end{align*}
for public input $(\mathbf{H}, \mathbf{A}, \widehat{\Gmc}^{(1)}, \widehat{\Gmc}^{(2)}, \mathbf{c}^{(1)}, \mathbf{c}^{(2)})$. 

\begin{figure*}[!htbp]
\begin{enumerate}
\item {\bf Commitment:} $\mathcal{P}$ samples the following uniformly random objects.
\begin{align*}
&\mathbf{b} \lr \{0,1\}^\ell, \quad 
\pi \lr \mathsf{S}_m, \quad \sigma^{(1)}, \sigma^{(2)}\lr \mathsf{S}_{\nmc}, \quad 
\rho_1, \rho_2, \rho_3 \lr \{0,1\}^\lambda, \\
&\mathbf{r}_{\mathbf{s}} \lr\F_{2}^{\ncfs}, \quad 
\mathbf{r}_{\mathbf{x}} \lr \F_2^N, \quad 
\mathbf{r}_{\mathbf{u}}^{(1)}, \mathbf{r}_{\mathbf{u}}^{(2)} \lr\F_{2}^{\kmc-\ell}, \quad 
\mathbf{r}_{\mathbf{f}}\lr\F_2^{2\ell}, \quad 
\mathbf{r}_{{\mathbf{e}}}^{(1)}, \mathbf{r}_{{\mathbf{e}}}^{(2)}\lr\F_{2}^{\nmc}.
\end{align*}
It then sends the commitment $\mathsf{CMT}:=(c_1, c_2, c_3)$ to $\mathcal{V}$, where
\begin{align*}
c_1 &= \text{COM}\left(\mathbf{b}, \hspace*{2.5pt}\pi,\hspace*{2.5pt} \{\sigma^{(i)}\}_{i \in [2]}, \hspace*{2.5pt} \mathbf{H}\cdot \mathbf{r}_{\mathbf{s}}^\top \hspace*{1.5pt}\oplus \hspace*{1.5pt}\mathbf{A}\cdot \mathbf{r}_{\mathbf{x}}^\top,\hspace*{2.5pt} \{(\hspace*{1.5pt}\mathbf{r}_{\mathbf{u}}^{(i)}\hspace*{1.5pt} \| \hspace*{1.5pt} \mathbf{r}_{\mathbf{f}}\hspace*{1.5pt})\cdot \widehat{\Gmc}^{(i)} \hspace*{1.5pt}\oplus\hspace*{1.5pt}  \mathbf{r}_{{\mathbf{e}}}^{(i)}\}_{i \in [2]}; \hspace*{5.5pt} \rho_1 \right),\\
c_2 &= \text{COM} \left(\pi(\mathbf{r}_{\mathbf{s}}), T_{\mathbf{b}}(\mathbf{r}_{{\mathbf{x}}}), \hspace*{2.55pt}T'_{\mathbf{b}}(\mathbf{r}_{\mathbf{f}}), \hspace*{2.55pt} \{\sigma^{(i)}(\mathbf{r}_{{\mathbf{e}}}^{(i)})\}_{i \in [2]}; \hspace*{5.5pt} \rho_2\right),\\
c_3 &= \text{COM}\left(\pi(\mathbf{s}\oplus \mathbf{r}_{\mathbf{s}}), \hspace*{2.55pt} T_{\mathbf{b}}({\mathbf{x}}\oplus\mathbf{r}_{{\mathbf{x}}}), \hspace*{2.55pt}T'_{\mathbf{b}}(\mathbf{f}\oplus\mathbf{r}_{\mathbf{f}}),\hspace*{2.55pt} \{\sigma^{(i)}({\mathbf{e}}^{(i)}\oplus\mathbf{r}_{{\mathbf{e}}}^{(i)})\}_{i \in [2]}; \hspace*{5.5pt} \rho_3\right).
\end{align*}

\item {\bf Challenge:} Upon receiving $\mathsf{CMT}$, $\mathcal{V}$ sends a challenge $\text{Ch}\lr\{1,2,3\}$ to $\mathcal{P}$.
\item {\bf Response:} $\mathcal{P}$ responds accordingly.
\begin{enumerate}
\item If $\text{Ch}=1$: Reveal $c_2$ and $c_3$. Let $ \mathbf{b}_1 = \itob(j)\oplus \mathbf{b}$,
\begin{equation*}
\mathbf{v}_{\mathbf{s}} = \pi(\mathbf{r}_{\mathbf{s}}), \quad
\mathbf{w}_{\mathbf{s}} = \pi({\mathbf{s}}),\quad
\mathbf{v}_{{\mathbf{x}}} = T_{\mathbf{b}}(\mathbf{r}_{{\mathbf{x}}}),\quad \mathbf{v}_{\mathbf{f}} = T'_{\mathbf{b}}(\mathbf{r}_{\mathbf{f}}),\quad 
\{\mathbf{v}_{{\mathbf{e}}}^{(i)} = \sigma^{(i)}(\mathbf{r}_{{\mathbf{e}}}^{(i)})\}_{i\in [2]}, \quad \{\mathbf{w}_{{\mathbf{e}}}^{(i)} = \sigma^{(i)}({\mathbf{e}}^{(i)})\}_{i \in [2]}.
\end{equation*}
Send $\mathsf{RSP}:=\left(\mathbf{b}_1, \hspace*{2.5pt} \mathbf{v}_{\mathbf{s}}, \hspace*{2.5pt} \mathbf{w}_{\mathbf{s}}, \hspace*{2.5pt}\mathbf{v}_{{\mathbf{x}}}, 			\hspace*{2.5pt}\mathbf{v}_{\mathbf{f}},
\hspace*{2.5pt}\{\mathbf{v}_{{\mathbf{e}}}^{(i)}, \hspace*{2.5pt}\mathbf{w}_{{\mathbf{e}}}^{(i)}\}_{i \in [2]}; \hspace*{2.55pt} \rho_2, \rho_3 \right)$ to $\mathcal{V}$.

\smallskip
			
\item If $\text{Ch}=2$: Reveal $c_1$ and $c_3$. Let 
\begin{align*}
&\mathbf{b}_2 = \mathbf{b}, \quad
\pi_2 = \pi, \quad
\{\sigma_2^{(i)} = \sigma^{(i)}\}_{i \in [2]}, \quad 
\mathbf{z}_{\mathbf{s}}=\mathbf{s}\oplus\mathbf{r}_{\mathbf{s}}, \quad \mathbf{z}_\mathbf{x}=\mathbf{x}\oplus\mathbf{r}_{\mathbf{x}}, \\
&\{\mathbf{z}_\mathbf{u}^{(i)}=\mathbf{u}^{(i)}\oplus\mathbf{r}_{\mathbf{u}}^{(i)}\}_{i \in [2]}, \quad \mathbf{z}_{\mathbf{f}}=\mathbf{f}\oplus\mathbf{r}_{\mathbf{f}}, \quad \{\mathbf{z}_{{\mathbf{e}}}^{(i)} = {\mathbf{e}}^{(i)}\oplus\mathbf{r}_{{\mathbf{e}}}^{(i)}\}_{i \in [2]}.
\end{align*}
			
Send $\mathsf{RSP}:=\left(\mathbf{b}_2, \hspace*{2.5pt}\pi_2, \hspace*{2.5pt}\{\sigma_2^{(i)}\}_{i \in [2]}, \hspace*{2.5pt}\mathbf{z}_{\mathbf{s}}, \hspace*{2.5pt}\mathbf{z}_{\mathbf{x}}, \hspace*{2.5pt}\{\mathbf{z}_{\mathbf{u}}^{(i)}\}_{i \in [2]}, \hspace*{2.5pt}\mathbf{z}_{\mathbf{f}}, \hspace*{2.5pt}\{\mathbf{z}_{{\mathbf{e}}}^{(i)}\}_{i \in [2]}; \hspace*{2.55pt} \rho_1, \rho_3\right)$ to $\mathcal{V}$.
			
\smallskip

\item If $\text{Ch}=3$: Reveal $c_1$ and $c_2$. Let $\mathbf{b}_3 = \mathbf{b}$,
\[
\pi_3 = \pi, \quad 
\{\sigma_3^{(i)} = \sigma^{(i)}\}_{i \in [2]}, \quad 
\mathbf{y}_{\mathbf{s}}=\mathbf{r}_{\mathbf{s}}, \quad \mathbf{y}_\mathbf{x}=\mathbf{r}_{\mathbf{x}}, \quad \{\mathbf{y}_\mathbf{u}^{(i)}=\mathbf{r}_{\mathbf{u}}^{(i)}\}_{i \in [2]}, \quad \mathbf{y}_{\mathbf{f}}=\mathbf{r}_{\mathbf{f}}, \quad \{\mathbf{y}_{{\mathbf{e}}}^{(i)} = \mathbf{r}_{{\mathbf{e}}}^{(i)}\}_{i \in [2]}.
\]

Send $\mathsf{RSP}:=\left(\mathbf{b}_3, \hspace*{2.5pt}\pi_3, \hspace*{1.5pt}\{\sigma_3^{(i)}\}_{i \in [2]}, \hspace*{1.5pt}\mathbf{y}_{\mathbf{s}}, \hspace*{2.5pt}\mathbf{y}_{\mathbf{x}}, \hspace*{2.5pt}\{\mathbf{y}_{\mathbf{u}}^{(i)}\}_{i \in [2]}, \hspace*{1.5pt}\mathbf{y}_{\mathbf{f}}, \hspace*{2.5pt}\{\mathbf{y}_{{\mathbf{e}}}^{(i)}\}_{i \in [2]}; \hspace*{2.55pt} \rho_1, \rho_2\right)$ to $\mathcal{V}$.
\end{enumerate}

\smallskip

\item {\bf Verification:} Upon receiving $\mathsf{RSP}$, $\mathcal{V}$ proceeds as follows.
\begin{enumerate}
\item If $\text{Ch}=1$: Let $\mathbf{w}_{\mathbf{x}} = \delta^N_{\btoi(\mathbf{b}_1)} \in \F_2^N$ and   $\mathbf{w}_{\mathbf{f}}=\Encode(\btoi(\mathbf{b}_1)) \in \F_2^{2\ell}$.\\
Check that ${\mathbf{w}}_{\mathbf{s}}\in \mathsf{B}(m, \omega)$ and $\{\mathbf{w}_{{\mathbf{e}}}^{(i)} \in \mathsf{B}(\nmc, \tmc)\}_{i \in [2]}$, and that
\begin{align*}
c_2 &= \text{COM}\left({\mathbf{v}}_{\mathbf{s}}, \hspace*{5pt}{\mathbf{v}}_{\mathbf{x}}, \hspace*{5pt}{\mathbf{v}}_{\mathbf{f}}, \hspace*{5pt}\{{\mathbf{v}}_{{\mathbf{e}}}^{(i)}\}_{i \in [2]}; \hspace*{5pt} \rho_2\right) \mbox{ and }\\
c_3 &= \text{COM}\left({\mathbf{v}}_{\mathbf{s}}\oplus{\mathbf{w}}_{\mathbf{s}}, \hspace*{5pt}{\mathbf{v}}_{\mathbf{x}}\oplus{\mathbf{w}}_{\mathbf{x}}, \hspace*{5pt}{\mathbf{v}}_{\mathbf{f}}\oplus{\mathbf{w}}_{\mathbf{f}}, \hspace*{5pt} \{{\mathbf{v}}_{{\mathbf{e}}}^{(i)}\oplus{\mathbf{w}}_{{\mathbf{e}}}^{(i)}\}_{i \in [2]}; \hspace*{5pt} \rho_3\right).
\end{align*}


\item If $\text{Ch}=2$: Check that
\begin{align*}
c_1 &= \text{COM}\left(\mathbf{b}_2,\hspace*{0.5pt} \pi_2, \hspace*{0.5pt}\{\sigma_2^{(i)}\}_{i \in [2]}, \hspace*{0.5pt}\Hcfs\cdot\mathbf{z}_{\mathbf{s}}^\top \oplus \mathbf{A} \cdot \mathbf{z}_{\mathbf{x}}^\top, \hspace*{0.5pt}\{(\mathbf{z}_{\mathbf{u}}^{(i)}\hspace*{0.5pt}\|\hspace*{0.5pt}\mathbf{z}_{\mathbf{f}}\hspace*{0.5pt})\cdot\widehat{\Gmc}^{(i)}\oplus\mathbf{z}_{{\mathbf{e}}}^{(i)} \oplus\mathbf{c}^{(i)}\}_{i \in [2]}; \hspace*{0.5pt} \rho_1\right),\\
c_3 &= \text{COM}
\left(
\pi_2(\mathbf{z}_{\mathbf{s}}), \hspace*{5pt} T_{\mathbf{b}_2}(\mathbf{z}_{{\mathbf{x}}}), \hspace*{5pt} T'_{\mathbf{b}_2}(\mathbf{z}_{\mathbf{f}}), \hspace*{5pt} \{\sigma_2^{(i)}(\mathbf{z}_{{\mathbf{e}}}^{(i)})\}_{i \in [2]}; \hspace*{5pt} \rho_3 
\right).
\end{align*}


\item If $\text{Ch}=3$: Check that
\begin{align*}
c_1 &= \text{COM}\left(\mathbf{b}_3,\hspace*{1.5pt} \pi_3, \hspace*{1.5pt}\{\sigma_3^{(i)}\}_{i \in [2]}, \hspace*{1.5pt}\Hcfs\cdot\mathbf{y}_{\mathbf{s}}^\top \hspace*{1.5pt}\oplus \hspace*{1.5pt}\mathbf{A}\cdot \mathbf{y}_{\mathbf{x}}^\top , \hspace*{1.5pt}\{(\hspace*{1.5pt}\mathbf{y}_{\mathbf{u}}^{(i)}\hspace*{1.5pt}\|\hspace*{1.5pt}\mathbf{y}_{\mathbf{f}}\hspace*{1.5pt})\cdot\widehat{\Gmc}^{(i)}\hspace*{1.5pt}\oplus\hspace*{1.5pt}\mathbf{y}_{{\mathbf{e}}}^{(i)}\}_{i \in [2]}; \hspace*{1.5pt} \rho_1\right),\\
c_2 &= \text{COM}\left(\pi_3(\mathbf{y}_{\mathbf{s}}), \hspace*{5pt} T_{\mathbf{b}_3}(\mathbf{y}_{{\mathbf{x}}}), \hspace*{5pt}T'_{\mathbf{b}_3}(\mathbf{y}_{\mathbf{f}}), \hspace*{5pt} \{\sigma_3^{(i)}(\mathbf{y}_{{\mathbf{e}}}^{(i)})\}_{i \in [2]}; \hspace*{5pt} \rho_2\right).
\end{align*}
\end{enumerate}

\smallskip

In each case, $\mathcal{V}$ outputs $1$ if and only if all the conditions hold. Otherwise, $\mathcal{V}$ outputs $0$.
\end{enumerate}	
\caption{The underlying ZK protocol of the CCA-anonymous group signature.}
\label{Figure:CCA-Protocol}
\end{figure*}

A ZK argument for the obtained equivalent statement can then be obtained in Stern's framework, using the same permuting and masking techniques of Section~\ref{sec:nizk}. The resulting interactive protocol is described in \textbf{Fig.}~\ref{Figure:CCA-Protocol}, where COM is a collision-resistant hash function modelled as a random oracle. The protocol is a statistical \textsf{ZKAoK}. Its simulator and extractor are constructed in the same manner as for the protocol in Section~\ref{sec:nizk}.  
The details can therefore be safely omitted here.

\section{A CPA-Anonymous Code-Based Group Signature}\label{sec:main-construction}
This section discusses our basic scheme that achieves \textsf{CPA}-anonimity. We start with a description of the scheme and end with the evaluation of its properties.
  
\subsection{Description, Efficiency, and Correctness of the Scheme} \label{subsec:Scheme-description}
Our group signature scheme consists of the following four algorithms.
\begin{enumerate}
\item \textsf{KeyGen}$(1^\lambda, 1^N)$: On input a security parameter $\lambda$ and an expected number of group users $N=2^{\ell} \in \mathsf{poly}(\lambda)$, for some positive integer $\ell$, this algorithm first selects the following parameters and hash functions.
\begin{enumerate}
	\item Parameters $\nmc = \nmc(\lambda), \kmc = \kmc(\lambda), \tmc = \tmc(\lambda)$ for a binary $[\nmc, \kmc, 2\tmc+1]$ Goppa code. 
	\item Parameters $m = m(\lambda), r= r(\lambda), \omega = \omega(\lambda)$ for the Syndrome Decoding problem, such that
	\begin{equation}\label{eqn:syndrome-decoding-parameters}
	r \leq \log\binom{m}{w} - 2\lambda - \mathcal{O}(1).
	\end{equation}	
	\item Two collision-resistant hash functions, to be modelled as random oracles:
	\begin{enumerate}
		\item $\mathrm{COM}: \{0,1\}^* \rightarrow \{0,1\}^\lambda$ to generate zero-knowledge arguments.
		\item $\hash:\{0,1\}^{*}\rightarrow \{1,2,3\}^\kappa$, where $\kappa = \omega \cdot \log{\lambda}$, for the Fiat-Shamir transformation.
	\end{enumerate}
\end{enumerate}

\smallskip

The algorithm then performs the following steps.
\begin{enumerate}
\item Run $\mckeygen(\nmc, \kmc, \tmc)$ to obtain a key pair   $\big(\pkmc=\Gmc\in\F_{2}^{\kmc\times\nmc}\hspace*{5pt}; \hspace*{5pt} \skmc\big)$ for the randomized McEliece encryption scheme with respect to a binary $[\nmc, \kmc, 2\tmc+1]$ Goppa code. The plaintext space is $\F_{2}^{\ell}$.
		
\item Choose a matrix $\Hcfs \xleftarrow{\$} \F_2^{r \times m}$.
\item For each $j \in [0, N\hspace*{-1pt}-\hspace*{-1pt}1]$, pick $\mathbf{s}_j \xleftarrow{\$} \mathsf{B}(m, \omega)$ and let $\y_j \in \F_2^{r}$ be its syndrome, \ie, $\y_j^\top=\Hcfs\cdot\mathbf{s}_j^{\top}$.
	
\smallskip
	
\item Output
	\begin{equation}\label{eqn:keygen-output}
		\hspace*{-15pt}\big(\mathsf{gpk} = (\Gmc, \mathbf{H}, \y_0, \ldots, \y_{N-1}), \hspace*{2.5pt}\mathsf{gmsk} = \skmc, \hspace*{2.5pt}\mathsf{gsk}= (\mathbf{s}_0, \ldots, \mathbf{s}_{N-1})\big).
	\end{equation}
\end{enumerate}
\item \textsf{Sign}$(\mathsf{gsk}[j], M)$: To sign a message $M \in \{0,1\}^*$ under $\mathsf{gpk}$, the group user of index $j$, who possesses secret key $\mathbf{s} = \mathsf{gsk}[j]$, performs the following steps.
\begin{enumerate}
\item Encrypt the binary representation of $j$, \ie, $\itob(j) \in \F_2^\ell$, under the randomized McEliece encrypting key $\mathbf{G}$. This is done by sampling 
$(\mathbf{u} \xleftarrow{\$} \F_2^{\kmc - \ell}, \mathbf{e} \xleftarrow{\$} \mathsf{B}(\nmc, \tmc))$ and outputting the ciphertext
$\mathbf{c} = \big(\hspace*{1.5pt}\mathbf{u}\hspace*{1.5pt}\|\hspace*{1.5pt}\itob(j)\hspace*{1.5pt}\big)\cdot\Gmc \oplus \mathbf{e} \in \F_2^{\nmc}$.
\item Generate a \textsf{NIZKAoK}  $\Pi$ to simultaneously prove, in zero-knowledge, the possession of an $\mathbf{s}\in \mathsf{B}(m, \omega)$ corresponding to a certain syndrome $\y_j \in \{\y_0, \ldots, \y_{N-1}\}$ with hidden index $j$, and that $\mathbf{c}$ is a correct McEliece encryption of $\itob(j)$. This is done by employing the interactive argument system in Section~\ref{sec:nizk} with public input $(\Gmc, \mathbf{H}, \y_0, \ldots, \y_{N-1}, \mathbf{c})$, and prover's witness $(j,\mathbf{s},\mathbf{u},\mathbf{e})$ that satisfies
\begin{equation}\label{eqn:relation-signing}
\mathbf{H}\cdot \mathbf{s}^\top = \y_j^\top,\quad 
\mathbf{s} \in \mathsf{B}(m, \omega), \quad 
\big(\hspace*{1.5pt}\mathbf{u}\hspace*{1.5pt} \| \hspace*{1.5pt}\itob(j)\hspace*{1.5pt}\big)\cdot \Gmc \oplus \mathbf{e} = \mathbf{c}, \quad  
\mathbf{e} \in \mathsf{B}(\nmc, \tmc).
\end{equation}
The protocol is repeated $\kappa=\omega \cdot \log{\lambda}$ times to achieve negligible soundness error, before being made non-interactive by using the Fiat-Shamir heuristic. We have
\begin{equation}\label{eqn:NIZKAoK-Pi}
\hspace*{-15pt}\Pi = \left(\CMT^{(1)}, \ldots, \CMT^{(\kappa)}; \hspace*{2.5pt}(\Ch^{(1)}, \ldots, \Ch^{(\kappa)});\hspace*{2.5pt}\RSP^{(1)}, \ldots, \RSP^{(\kappa)}\right),
\end{equation}
where $\left(\Ch^{(1)}\hspace*{-1pt}, \ldots, \Ch^{(\kappa)}\right) 
= \hash\left(M; \CMT^{(1)}\hspace*{-1pt}, \ldots, \CMT^{(\kappa)}; \mathsf{gpk}, \mathbf{c}\right) \in \{1,2,3\}^\kappa$. 
\item Output the group signature $\Sigma=(\mathbf{c}, \Pi)$.
\end{enumerate}

\smallskip
	
\item \textsf{Verify}$(\mathsf{gpk}, M, \Sigma)$: Parse $\Sigma$ as $(\mathbf{c}, \Pi)$, parse $\Pi$ as in~(\ref{eqn:NIZKAoK-Pi}), and then proceed as follows.

\begin{enumerate}
\item If $\left(\Ch^{(1)}, \ldots, \Ch^{(\kappa)}\right) \neq 
\hash\left(M; \CMT^{(1)}, \ldots, \CMT^{(\kappa)}; \mathsf{gpk}, \mathbf{c}\right)$, then return $0$.
\item For $i=1$ to $\kappa$, run the verification step of the interactive protocol in Section~\ref{sec:nizk} on public input $(\Gmc, \mathbf{H}, \y_0, \ldots, \y_{N-1}, \mathbf{c})$ to check the validity of $\RSP^{(i)}$ with respect to $\CMT^{(i)}$ and $\Ch^{(i)}$. If any of the verification conditions fails to hold, then return $0$.
\item Return $1$.
\end{enumerate}
\item \textsf{Open}$(\mathsf{gmsk}, M, \Sigma)$: Parse $\Sigma$ as $(\mathbf{c}, \Pi)$ and run $\mcdec(\mathsf{gmsk},\mathbf{c})$ to decrypt $\mathbf{c}$. If decryption fails, then return $\bot$. If decryption outputs $\mathbf{g} \in \F_2^\ell$, then return $j = \btoi(\mathbf{g}) \in [0,N-1]$.
\end{enumerate}

\begin{remark}\label{remark:uniform-syndrome} 
Lemma~\ref{lemma:Leftover-Hash-Lemma} assures us that, for the parameters $m, r, \omega$ that satisfy the inequality condition in~(\ref{eqn:syndrome-decoding-parameters}), the distribution of syndrome $\mathbf{y}_j$, for all $j \in [0, N\hspace*{-1pt}-\hspace*{-1pt}1]$, is statistically close to the uniform distribution over $\F_2^r$.
\end{remark}

The efficiency, correctness, and security aspects of the above group signature scheme can now be summarized into the following theorem.
\begin{theorem}
The given group signature scheme is correct. The public key has size $nk + (m+N)r$ bits. The bit-size of the signatures is bounded above by $\big((N + 3\log N) + m(\log m +1) + n(\log n +1) + k + 5\lambda\big)\kappa +n$. In the random oracle model we can make two further assertions. First, if the Decisional McEliece problem $\mathsf{DMcE}(\nmc, \kmc, \tmc)$ and the Decisional Learning Parity with fixed-weight Noise problem $\mathsf{DLPN}(\kmc-\ell, \nmc, \mathsf{B}(\nmc, \tmc))$ are hard, then the scheme is $\mathsf{CPA}$-anonymous. Second, if the Syndrome Decoding problem $\mathsf{SD}(m, r, \omega)$ is hard, then the scheme is traceable.
\end{theorem}

In terms of efficiency, it is clear from (\ref{eqn:keygen-output}) that $\mathsf{gpk}$ has bit-size $nk + (m+N)r$. The length of the \textsf{NIZKAoK} $\Pi$ is $\kappa$ times the communication cost of the underlying interactive protocol. Thus, by Theorem~\ref{theorem:Protocol-properties}, $\Sigma=(\mathbf{c}, \Pi)$ has bit-size bounded above by $\big((N + 3\log N) + m(\log m +1) + n(\log n +1) + k + 5\lambda\big)\kappa +n$.

To see that the given group signature scheme is correct, first observe that the honest user with index $j$, for any $j \in [0, N-1]$, can always obtain a tuple $(j,\mathbf{s},\mathbf{u},\mathbf{e})$ satisfying~(\ref{eqn:relation-signing}). Then, since the underlying interactive protocol is perfectly complete, $\Pi$ is a valid \textsf{NIZKAoK} and algorithm $\mathsf{Verify}(\mathsf{gpk}, M, \Sigma)$ always outputs $1$, for any message $M \in \{0,1\}^*$. On the correctness of \textsf{Open}, it suffices to note that, if the ciphertext $\mathbf{c}$ is of the form $\mathbf{c} = \big(\hspace*{1.5pt}\mathbf{u}\hspace*{1.5pt}\|\hspace*{1.5pt}\itob(j)\hspace*{1.5pt}\big)\cdot\Gmc \oplus \mathbf{e}$, where $\mathbf{e} \in \mathsf{B}(\nmc, \tmc)$, then, by the correctness of the randomized McEliece encryption scheme, $\mcdec(\mathsf{gmsk},\mathbf{c})$ outputs $\itob(j)$.

\subsection{Anonymity}\label{subsec:proof-anonymity}
Let $\mathcal{A}$ be any PPT adversary attacking the \textsf{CPA}-anonymity of the scheme with advantage $\epsilon$. We will prove that $\epsilon=\mathsf{negl}(\lambda)$ based on the \textsf{ZK} property of the underlying argument system. To do so, we retain the assumed hardness of the $\mathsf{DMcE}(\nmc, \kmc, \tmc)$ and the $\mathsf{DLPN}(\kmc-\ell, \nmc, \mathsf{B}(\nmc, \tmc))$ problems. Specifically, we consider the following sequence of hybrid experiments $G_0^{(b)}$, $G_1^{(b)}$, $G_2^{(b)}$, $G_3^{(b)}$, and $G_4$.

\smallskip

\noindent
\textbf{Experiment} $G_0^{(b)}$. This is the real \textsf{CPA}-anonymity game. The challenger runs $\mathsf{KeyGen}(1^\lambda, 1^N)$ to obtain 
\begin{equation*}
	\big(\hspace*{2.5pt}\mathsf{gpk} = (\Gmc, \mathbf{H}, \y_0, \ldots, \y_{N-1}), \hspace*{5pt}\mathsf{gmsk} = \skmc, \hspace*{5pt}\mathsf{gsk}= (\mathsf{gsk}[0], \ldots, \mathsf{gsk}[N-1])\hspace*{2.5pt}\big),
\end{equation*}
and then gives $\mathsf{gpk}$ and $\{\mathsf{gsk}[j]\}_{j \in [0, N-1]}$ to $\mathcal{A}$.  In the challenge phase, $\mathcal{A}$ outputs a message $M^*$ together with two indices $j_0, j_1 \in [0, N-1]$. The challenger sends back a challenge signature $\Sigma^* = (\mathbf{c}^*, \Pi^*) \leftarrow \mathsf{Sign}(\mathsf{gpk}, \mathsf{gsk}[j_b])$, where $\mathbf{c}^* = \big(\hspace*{1.5pt}\mathbf{u}\hspace*{1.5pt}\|\hspace*{1.5pt}\itob(j_b)\hspace*{1.5pt}\big)\cdot\Gmc \oplus \mathbf{e}$, with $\mathbf{u} \xleftarrow{\$} \F_2^{\kmc - \ell}$ and $\mathbf{e} \xleftarrow{\$} \mathsf{B}(\nmc, \tmc)$. The adversary then outputs $b$ with probability $1/2 + \epsilon$.

\smallskip

\noindent
\textbf{Experiment} $G_1^{(b)}$. This experiment introduces a modification in the challenge phase. Instead of faithfully generating the \textsf{NIZKAoK} $\Pi^*$, the challenger simulates it as follows.
\begin{enumerate}
\item Compute $\mathbf{c}^* \in \F_2^{\nmc}$ as in Experiment $G_0^{(b)}$.
	\item Run the simulator of the underlying interactive protocol 
	$\kappa= \omega(\log \lambda)$ times on input $(\Gmc, \mathbf{H}, \y_0, \ldots, \y_{N-1}, \mathbf{c}^*)$. Program the random oracle $\hash$ accordingly.
	\item Output the simulated \textsf{NIZKAoK} $\Pi^*$.
\end{enumerate}
Since the underlying argument system is statistically zero-knowledge, $\Pi^*$ is statistically close to the real \textsf{NIZKAoK}. As a result, the simulated signature $\Sigma^* = \big(\mathbf{c}^*, \Pi^*\big)$ is statistically close to the one in experiment $G_0^{(b)}$. It then follows that $G_0^{(b)}$ and $G_1^{(b)}$ are indistinguishable from $\mathcal{A}$'s view.

\smallskip

\noindent
\textbf{Experiment} $G_2^{(b)}$. This experiment makes the following change with respect to $G_1^{(b)}$. The encrypting key $\Gmc$ obtained from $\mckeygen(\nmc, \kmc, \tmc)$ is replaced by a uniformly random matrix $\Gmc \xleftarrow{\$} \F_2^{\kmc\times\nmc}$. Lemma~\ref{Lemma3} will demonstrate that Experiments $G_1^{(b)}$ and $G_2^{(b)}$ are computationally indistinguishable
based on the assumed hardness of the $\mathsf{DMcE}(\nmc, \kmc, \tmc)$ problem.

\begin{lemma}\label{Lemma3}
If $\mathcal{A}$ can distinguish Experiments $G_1^{(b)}$ and $G_2^{(b)}$ with probability non-negligibly larger than $1/2$, then there exists an efficient distinguisher $\mathcal{D}_1$ that solves the $\mathsf{DMcE}(\nmc, \kmc, \tmc)$ problem with the same probability.
\end{lemma}

\begin{IEEEproof}
An instance of the $\mathsf{DMcE}(\nmc, \kmc, \tmc)$ problem is $\Gmc^* \in \F_2^{\kmc\times\nmc}$, which can either be uniformly random or be generated by 
$\mckeygen(\nmc, \kmc, \tmc)$. The distinguisher $\mathcal{D}_1$ receives a challenge instance $\Gmc^*$ and uses $\mathcal{A}$ to distinguish between the two. It interacts with $\mathcal{A}$ by performing the following steps.
	
\begin{enumerate}
\item {\sf Setup. } Generate $(\mathbf{H}, \y_0, \ldots, \y_{N-1})$ and $(\mathsf{gsk}[0], \ldots, \mathsf{gsk}[N-1])$ as in the real scheme. Send $\mathcal{A}$ the pair
\begin{equation*}
\big(\hspace*{2.5pt}\mathsf{gpk}^* = (\Gmc^*, \mathbf{H}, \y_0, \ldots, \y_{N-1}), \hspace*{5pt}\mathsf{gsk}= (\mathsf{gsk}[0], \ldots, \mathsf{gsk}[N-1])\hspace*{2.5pt}\big).
\end{equation*}
		
\item {\sf Challenge.} Upon receiving the challenge $(M^*, j_0, j_1)$, $\mathcal{D}_1$ proceeds as follows.
\begin{enumerate}
\item Pick $b \xleftarrow{\$} \{0,1\}$, and compute $\mathbf{c}^* = \big(\hspace*{1.5pt}\mathbf{u}\hspace*{1.5pt}\|\hspace*{1.5pt}\itob(j_b)\hspace*{1.5pt}\big)\cdot\Gmc^* \oplus \mathbf{e}$, where $\mathbf{u} \xleftarrow{\$} \F_2^{\kmc - \ell}$ and $\mathbf{e} \xleftarrow{\$} \mathsf{B}(\nmc, \tmc)$.
\item Simulate the \textsf{NIZKAoK} $\Pi^*$ on input $(\Gmc^*, \mathbf{H}, \y_0, \ldots, \y_{N-1}, \mathbf{c}^*)$, and output $\Sigma^* = \big(\mathbf{c}^*, \Pi^*\big)$.
\end{enumerate}
\end{enumerate}
We observe that if $\Gmc^*$ is generated by $\mckeygen(\nmc, \kmc, \tmc)$ then the view of $\mathcal{A}$ in the interaction with $\mathcal{D}_1$ is statistically close to its view in Experiment $G_1^{(b)}$ with the challenger. On the other hand, if  $\Gmc^*$ is uniformly random, then $\mathcal{A}$'s view is statistically close to its view in Experiment $G_2^{(b)}$. Therefore, if $\mathcal{A}$ can guess whether it is interacting with the challenger in $G_1^{(b)}$ or $G_2^{(b)}$ with probability non-negligibly larger than $1/2$, then $\mathcal{D}_1$ can use $\mathcal{A}$'s guess to solve the challenge instance $\Gmc^*$ of the $\mathsf{DMcE}(\nmc, \kmc, \tmc)$ problem, with the same probability.
\end{IEEEproof}

\smallskip

\noindent
\textbf{Experiment} $G_3^{(b)}$. In Experiment $G_2^{(b)}$ we have 
\begin{equation*}
\mathbf{c}^* = \big(\hspace*{1.5pt}\mathbf{u}\hspace*{1.5pt}\|\hspace*{1.5pt}\itob(j_b)\hspace*{1.5pt}\big)\cdot\Gmc \oplus \mathbf{e} = (\mathbf{u}\cdot \Gmc_1 \oplus \mathbf{e}) \oplus \itob(j_b)\cdot\Gmc_2, 
\end{equation*}
where $\mathbf{u} \xleftarrow{\$} \F_2^{\kmc - \ell}$,  $\mathbf{e} \xleftarrow{\$} \mathsf{B}(\nmc, \tmc)$, $\Gmc_1 \in \F_2^{(\kmc - \ell) \times \nmc}$, and $\Gmc_2 \in \F_2^{\ell \times \nmc}$ such that $\Big[\frac{\mathbf{G}_1}{\mathbf{G}_2}\Big]= \Gmc$. Now, Experiment $G_3^{(b)}$ modifies the generation of $\mathbf{c}^*$ by replacing the original one with  $\mathbf{c}^* = \mathbf{v} \oplus \itob(j_b)\cdot\Gmc_2$, where $\mathbf{v} \xleftarrow{\$} \F_2^{\nmc}$. Experiments $G_2^{(b)}$ and $G_3^{(b)}$ are computationally indistinguishable based on the assumed hardness of the $\mathsf{DLPN}(\kmc-\ell, \nmc, \mathsf{B}(\nmc, \tmc))$ problem, as shown in Lemma~\ref{Lemma4}. 

\begin{lemma}\label{Lemma4}
If $\mathcal{A}$ can distinguish Experiments $G_2^{(b)}$ and $G_3^{(b)}$ with probability non-negligibly larger than $1/2$, then there exists an efficient distinguisher $\mathcal{D}_2$ solving the $\mathsf{DLPN}(\kmc-\ell, \nmc, \mathsf{B}(\nmc, \tmc))$ problem with the same probability.
\end{lemma}

\begin{IEEEproof}
An instance of the $\mathsf{DLPN}(\kmc-\ell, \nmc, \mathsf{B}(\nmc, \tmc))$ problem is a pair $(\mathbf{B}, \mathbf{v}) \in \F_2^{(\kmc - \ell) \times \nmc} \times \F_2^{\nmc}$, where $\mathbf{B}$ is uniformly random, and $\mathbf{v}$ is either uniformly random or of the form $\mathbf{v} = \mathbf{u} \cdot \mathbf{B} \oplus \mathbf{e}$, for $\mathbf{u} \xleftarrow{\$} \F_2^{\kmc - \ell}$ and $\mathbf{e} \xleftarrow{\$} \mathsf{B}(\nmc, \tmc)$. The distinguisher $\mathcal{D}_2$ receives a challenge instance $(\mathbf{B}, \mathbf{v})$ and uses $\mathcal{A}$ to distinguish between the two. It interacts with $\mathcal{A}$ in the following manners.
\begin{enumerate}
\item {\sf Setup. } Pick $\Gmc_2 \xleftarrow{\$} \F_2^{\ell \times \nmc}$ and let $\Gmc^* = \big[\frac{\mathbf{B}}{\Gmc_2}\big]$. Generate $(\mathbf{H}, \y_0, \ldots, \y_{N-1})$ and $(\mathsf{gsk}[0], \ldots, \mathsf{gsk}[N-1])$ as in the real scheme, and send $\mathcal{A}$ the pair $\big(\hspace*{2.5pt}\mathsf{gpk}^* = (\Gmc^*, \mathbf{H}, \y_0, \ldots, \y_{N-1}), \hspace*{5pt}\mathsf{gsk}= (\mathsf{gsk}[0], \ldots, \mathsf{gsk}[N-1])\hspace*{2.5pt}\big)$.
\item {\sf Challenge.} Upon receiving the challenge $(M^*, j_0, j_1)$, $\mathcal{D}_2$ first picks $b \xleftarrow{\$} \{0,1\}$ and lets $\mathbf{c}^* = \mathbf{v} \oplus \itob(j_b)\cdot\Gmc_2$, where $\mathbf{v}$ comes from the challenge \textsf{DLPN} instance. It then simulates the \textsf{NIZKAoK} $\Pi^*$, on input $(\Gmc^*, \mathbf{H}, \y_0, \ldots, \y_{N-1}, \mathbf{c}^*)$, and outputs $\Sigma^* = \big(\mathbf{c}^*, \Pi^*\big)$.
\end{enumerate}

If $\mathcal{D}_2$'s input pair $(\mathbf{B}, \mathbf{v})$ is of the form $(\mathbf{B}, \mathbf{v}= \mathbf{u} \cdot \mathbf{B} \oplus \mathbf{e})$, where $\mathbf{u} \xleftarrow{\$} \F_2^{\kmc - \ell}$ and $\mathbf{e} \lr \mathsf{B}(\nmc, \tmc)$, then the view of $\mathcal{A}$ in the interaction with $\mathcal{D}_2$ is statistically close to its view in Experiment $G_2^{(b)}$ with the challenger. On the other hand, if the pair $(\mathbf{B}, \mathbf{v})$ is uniformly random, then $\mathcal{A}$'s view is statistically close to its view in Experiment $G_3^{(b)}$. Therefore, if $\mathcal{A}$ can guess whether it is interacting with the challenger in $G_2^{(b)}$ or $G_3^{(b)}$ with probability non-negligibly larger than $1/2$, then $\mathcal{D}_2$ can use $\mathcal{A}$'s guess to solve the challenge instance of the $\mathsf{DLPN}(\kmc-\ell, \mathsf{B}(\nmc, \tmc))$ problem with the same probability.
\end{IEEEproof}

\smallskip 

\noindent
\textbf{Experiment} $G_4$. This experiment is a modification of Experiment $G_3^{(b)}$. The ciphertext $\mathbf{c}^*$ is now set as $\mathbf{c}^* = \mathbf{r} \xleftarrow{\$} \F_2^{\nmc}$. Clearly, the distributions of $\mathbf{c}^*$ in Experiments $G_3^{(b)}$ and $G_4$ are identical. As a result, $G_4$ and $G_3^{(b)}$ are statistically indistinguishable. We note that $G_4$ no longer depends on the challenger's bit $b$, and thus, $\mathcal{A}$'s advantage in this experiment is $0$.

The above discussion shows that Experiments $G_0^{(b)}, G_1^{(b)}, G_2^{(b)}, G_3^{(b)}, G_4$ are indistinguishable and that  $\mathbf{Adv}_{\mathcal{A}}(G_4)= 0$.
It then follows that the advantage of $\mathcal{A}$ in attacking the \textsf{CPA}-anonymity of the scheme, {{\ie}}, in experiment $G_0^{(b)}$, is negligible. The \textsf{CPA}-anonymity property is, thus, confirmed.

\subsection{Traceability}
Let $\mathcal{A}$ be a PPT traceability adversary against our group signature scheme with success probability $\epsilon$. We construct a PPT algorithm $\mathcal{F}$ that solves the $\mathsf{SD}(\ncfs, \kcfs, \tcfs)$ problem with success probability polynomially related to $\epsilon$.

Algorithm $\mathcal{F}$ receives a challenge $\mathsf{SD}(\ncfs, \kcfs, \tcfs)$ instance, \ie, a uniformly random matrix-syndrome pair $(\widetilde{\mathbf{H}}, \tilde{\mathbf{y}}) \in \F_2^{\kcfs \times \ncfs} \times \F_2^{\kcfs}$. The goal of $\mathcal{F}$ is to find a vector $\mathbf{s} \in \mathsf{B}(m, \omega)$ such that $\widetilde{\mathbf{H}}\cdot \mathbf{s}^\top = \tilde{\mathbf{y}}^\top$.
It then carry out the following tasks.
\begin{enumerate}
	\item Pick a guess $j^* \xleftarrow{\$} [0,N-1]$ and set $\mathbf{y}_{j^*} = \tilde{\mathbf{y}}$.
	\item Set $\mathbf{H} = \widetilde{\mathbf{H}}$. For each $j \in [0, N-1]$ such that $j \neq j^*$, sample $\mathbf{s}_j \lr \mathsf{B}(m,\omega)$ and set $\mathbf{y}_j\in \F_2^r$ to be its syndrome, \ie, $\mathbf{y}_j^\top = \mathbf{H}\cdot \mathbf{s}_j^\top $.
	\item Run $\mckeygen(\nmc, \kmc, \tmc)$ to obtain a key pair $\big(\pkmc=\Gmc\in\F_{2}^{\kmc\times\nmc}\hspace*{5pt}; \hspace*{5pt} \skmc\big)$.
	\item Send $\mathsf{gpk} = \big( \Gmc, \mathbf{H}, \mathbf{y}_0, \ldots, \mathbf{y}_{N-1}\big)$ and $\mathsf{gmsk} = \skmc$ to $\mathcal{A}$.
\end{enumerate}

Since the parameters $m, r, \omega$ were chosen such that $r \leq \log\binom{m}{w} - 2\lambda - \mathcal{O}(1)$, Lemma~\ref{lemma:Leftover-Hash-Lemma} says that the distribution of the syndrome $\mathbf{y}_j$, for all $j \neq j^*$, is statistically close to the uniform distribution over $\F_2^r$. In addition, the syndrome $\mathbf{y}_{j^*} = \tilde{\mathbf{y}}$ is truly uniform over $\F_2^r$. It then follows that the distribution of $(\mathbf{y}_0, \ldots, \mathbf{y}_{N-1})$ is statistically close to that of the real scheme as noted in Remark~\ref{remark:uniform-syndrome}.
As a result, the distribution of $(\mathsf{gpk}, \mathsf{gmsk})$ is statistically close to the distribution expected by $\mathcal{A}$.

The forger $\mathcal{F}$ then initializes a set $CU = \emptyset$ and handles the queries from $\mathcal{A}$ according to the following procedure.
\begin{enumerate}
	\item Queries to the random oracle $\mathcal{H}$ are handled by consistently returning uniformly random values in $\{1,2,3\}^\kappa$. Suppose that $\mathcal{A}$ makes $Q_{\mathcal{H}}$ queries, then for each $\eta \leq Q_{\mathcal{H}}$, we let $r_\eta$ denote the answer to the $\eta$-th query. \medskip
	\item $\mathcal{O}^\mathsf{Corrupt}(j)$, for any $j \in [0,N-1]$, depends on how $j$ and $j^*$ are related. If $j \neq j^*$, then $\mathcal{F}$ sets $CU: = CU \cup \{j\}$ and gives $\mathbf{s}_j$ to $\mathcal{A}$. If $j = j^*$, then $\mathcal{F}$ aborts.
	\item $\mathcal{O}^\mathsf{Sign}(j, M)$, for any $j \in [0,N-1]$ and any message $M$, also depends on $j$ and $j^*$. If $j \neq j^*$, then $\mathcal{F}$ honestly computes a signature, since it has the secret key $\mathbf{s}_j$. If $j = j^*$, then $\mathcal{F}$ returns a simulated signature $\Sigma^*$ computed as in Section~\ref{subsec:proof-anonymity}. Please consult, specifically, Experiment $G_1^{(b)}$ in the proof of anonymity.
\end{enumerate}

At some point, $\mathcal{A}$ outputs a forged group signature $\Sigma^*$ on some message $M^*$, where 
\[
\Sigma^* = \big(\mathbf{c}^*, \big(\CMT^{(1)}, \ldots, \CMT^{(\kappa)}; \hspace*{2.5pt}\Ch^{(1)}, \ldots, \Ch^{(\kappa)};\hspace*{2.5pt}\RSP^{(1)}, \ldots, \RSP^{(\kappa)}  \big) \big).
\]
By the requirements of the traceability experiment, one has $\mathsf{Verify}(\mathsf{gpk}, M^*, \Sigma^*)=1$ and, for all $j \in CU$, signatures of user $j$ on $M^*$ were never queried. Now $\mathcal{F}$ uses $\skmc$ to open $\Sigma^*$, and aborts if the opening algorithm does not output $j^*$. It can be checked that $\mathcal{F}$ aborts with probability at most 
${(N-1)}/{N} + (2/3)^\kappa$. This is because the choice of $j^* \in [0, N-1]$ is completely hidden from $\mathcal{A}$'s view and $\mathcal{A}$ can only violate the soundness of the argument system with probability at most $(2/3)^\kappa$. Thus, with probability at least $1/N - (2/3)^\kappa$, 
\begin{equation}\label{eqn:before-exploiting-the-forgery}
\mathsf{Verify}(\mathsf{gpk}, M^*, \Sigma^*)=1 \hspace*{5pt}\wedge \hspace*{5pt} \mathsf{Open}(\skmc, M^*, \Sigma^*) = j^*.
\end{equation}

Suppose that~(\ref{eqn:before-exploiting-the-forgery}) holds. Algorithm $\mathcal{F}$ then exploits the forgery as follows. Denote by $\Delta$ the tuple 
\[
\big(M^*; \CMT^{(1)}, \ldots, \CMT^{(\kappa)}; \Gmc, \mathbf{H}, \mathbf{y}_0, \ldots, \mathbf{y}_{N-1}, \mathbf{c}^*\big).
\]
Observe that if $\mathcal{A}$ has never queried the random oracle $\mathcal{H}$ on input $\Delta$, then $\mathrm{Pr}\big[\big(\Ch^{(1)}, \ldots, \Ch^{(\kappa)}\big) = \hash(\Delta)\big] \leq 3^{-\kappa}$.
Thus, with probability at least $\epsilon - 3^{-\kappa}$, there exists certain $\eta^* \leq Q_{\mathcal{H}}$ such that $\Delta$ was the input of the $\eta^*$-th query.
Next, $\mathcal{F}$ picks $\eta^*$ as the target forking point and replays $\mathcal{A}$ many times with the same random tape and input as in the original run. In each rerun, for the first $\eta^* -1$ queries, $\mathcal{A}$ is given the same answers $r_1, \ldots, r_{\eta^*-1}$ as in the initial run. From the $\eta^*$-th query onwards, however, $\mathcal{F}$ replies with fresh random values $r^{'}_{\eta^*}, \ldots, r^{'}_{q_{\mathcal{H}}} \xleftarrow{\$} \{1,2,3\}^\kappa$. The Improved Forking Lemma of Pointcheval and Vaudenay \cite[Lemma~7]{PV97} implies that, with probability larger than $1/2$ and within less than $32\cdot Q_{\mathcal{H}}/(\epsilon - 3^{-\kappa})$ executions of $\mathcal{A}$, algorithm $\mathcal{F}$ can obtain a $3$-fork involving the tuple $\Delta$. Now, let the answers of $\mathcal{F}$ with respect to the $3$-fork branches be
\[
r_{1,\eta^*}=(\mathsf{Ch}^{(1)}_1, \ldots, \mathsf{Ch}^{(\kappa)}_1); \hspace*{4pt} r_{2,\eta^*}=(\mathsf{Ch}^{(1)}_2, \ldots, \mathsf{Ch}^{(\kappa)}_2); \hspace*{4pt} r_{3,\eta^*}=(\mathsf{Ch}^{(1)}_3, \ldots, \mathsf{Ch}^{(\kappa)}_3).
\]
Then, by a simple calculation, one has
\[
\mathrm{Pr}\big[\hspace*{1.5pt}\exists \hspace*{1pt}i \in \{1, \ldots, \kappa\}: \hspace*{3pt} \{\mathsf{Ch}^{(i)}_1, \mathsf{Ch}^{(i)}_2, \mathsf{Ch}^{(i)}_3\} = \{1,2,3\}\big]= 1- (7/9)^\kappa.
\]

Conditioned on the existence of such index $i$, one parses the three forgeries corresponding to the fork branches to obtain $\big(\mathsf{RSP}^{(i)}_1, \mathsf{RSP}^{(i)}_2, \mathsf{RSP}^{(i)}_3\big)$. They turn out to be three \emph{valid} responses with respect to $3$ different challenges for the same commitment $\mathsf{CMT}^{(i)}$. Then, by using the knowledge extractor of the underlying interactive argument system (see Lemma~\ref{lemma:proof-of-knowledge-extraction}), one can efficiently extract
a tuple $(j', \mathbf{s}', \mathbf{u}', \mathbf{e}')\in [0, N-1] \times \F_2^{\ncfs} \times \F_2^{\kmc - \ell} \times \F_2^{\nmc}$ such that
\[
\mathbf{H}\cdot \mathbf{s}'^\top = \y_{j'}^\top, \quad 
\mathbf{s}'\in\mathsf{B}(m,\omega), \quad 
\big(\hspace*{1.5pt}\mathbf{u}'\hspace*{1.5pt} \| \hspace*{1.5pt}\itob(j')\hspace*{1.5pt}\big)\cdot \Gmc \hspace*{1.5pt}\oplus \hspace*{1.5pt}\mathbf{e}' = \mathbf{c}^*, \quad 
\mathbf{e}'\in \mathsf{B}(\nmc, \tmc).
\]
Since the given group signature scheme is correct, the equation
$\big(\hspace*{1.5pt}\mathbf{u}'\hspace*{1.5pt} \| \hspace*{1.5pt}\itob(j')\hspace*{1.5pt}\big)\cdot \Gmc \oplus \mathbf{e}' = \mathbf{c}^*$ implies that $\mathsf{Open}(\skmc, M^*, \Sigma^*) = j'$. On the other hand, we have $\mathsf{Open}(\skmc, M^*, \Sigma^*) = j^*$, which leads to $j' = j^*$. Therefore, it holds that $\widetilde{\mathbf{H}}\cdot\mathbf{s}'^{\top}= \mathbf{H}\cdot\mathbf{s}'^{\top} = \mathbf{y}_{j^*}^\top = \tilde{\mathbf{y}}^\top$ and $\mathbf{s}' \in \mathsf{B}(m, \omega)$. In other words, $\mathbf{s}'$ is a valid solution to the challenge $\mathsf{SD}(m, r, \omega)$ instance $(\widetilde{\mathbf{H}}, \tilde{\mathbf{y}})$.

Finally, the above analysis shows that, if $\mathcal{A}$ has success probability $\epsilon$ and running time $T$ in attacking the traceability of our group signature scheme, then $\mathcal{F}$ has success probability at least ${1}/{2}\big(1/N - (2/3)^\kappa\big)\big(1 - (7/9)^\kappa\big)$ and running time at most $32\cdot T\cdot Q_{\mathcal{H}}/(\epsilon - 3^{-\kappa}) + \mathsf{poly}(\lambda, N)$. This concludes the proof of the traceability property.

\section{Achieving CCA-Anonymity}\label{sec:CCA}

In this section, we propose and analyse a code-based group signature that achieves the strong notion of \textsf{CCA}-anonymity. The scheme is an extension of the \textsf{CPA}-anonymous scheme described in Section~\ref{subsec:Scheme-description}. To achieve \textsf{CCA}-security for the underlying encryption layer via the Naor-Yung transformation~\cite{NY90}, the binary representation of the signer's index $j$ is now verifiably encrypted twice under two different randomized McEliece public keys $\pkmc^{(1)}$ and $\pkmc^{(2)}$. This enables \textsf{CCA}-anonymity for the resulting group signature scheme. In describing the scheme, the focus is on presenting the modifications that we must make with respect to the earlier scheme from Section~\ref{subsec:Scheme-description}. 

\subsection{Description of the Scheme}\label{subsec:CCA-scheme-description}

We start with the four algorithms that constitute the scheme.

\begin{enumerate}
\item \textsf{KeyGen}$(1^\lambda, 1^N)$: The algorithm proceeds as the key generation algorithm of Section~\ref{subsec:Scheme-description}, with the following alteration. $\mckeygen(\nmc, \kmc, \tmc)$ is run twice, producing two  key pairs $(\pkmc^{(1)}=\Gmc^{(1)}, \skmc^{(1)})$ and $(\pkmc^{(2)}=\Gmc^{(2)}, \skmc^{(2)})$ for the randomized McEliece encryption. Then $\Gmc^{(1)}$ and $\Gmc^{(2)}$ are included in the group public key $\mathsf{gpk}$, the opening secret key $\mathsf{gmsk}$ is defined to be $\skmc^{(1)}$, while $\skmc^{(2)}$ is discarded. 

\smallskip
	
\item \textsf{Sign}$(\mathsf{gsk}[j], M)$: The binary representation $\itob(j) \in \F_2^\ell$ of the user's index $j$ is now encrypted twice, under the keys $\Gmc^{(1)}$ and $\Gmc^{(2)}$. The resulting ciphertexts $\mathbf{c}^{(1)}$ and $\mathbf{c}^{(2)}$ have the form
\[
\mathbf{c}^{(1)} = \big(\hspace*{1.5pt}\mathbf{u}^{(1)}\hspace*{1.5pt}\|\hspace*{1.5pt}\itob(j)\hspace*{1.5pt}\big)\cdot\Gmc^{(1)} \oplus \mathbf{e}^{(1)} \in \F_2^{\nmc} ~~\mbox{ and }~~
\mathbf{c}^{(2)} = \big(\hspace*{1.5pt}\mathbf{u}^{(2)}\hspace*{1.5pt}\|\hspace*{1.5pt}\itob(j)\hspace*{1.5pt}\big)\cdot\Gmc^{(2)} \oplus \mathbf{e}^{(2)} \in \F_2^{\nmc}, 
\]
where $\mathbf{u}^{(1)}, \mathbf{u}^{(2)} \xleftarrow{\$} \F_2^{\kmc - \ell}$ and $\mathbf{e}^{(1)}, \mathbf{e}^{(2)} \xleftarrow{\$} \mathsf{B}(\nmc, \tmc)$. 
	
The zero-knowledge protocol of the scheme from Section~\ref{subsec:Scheme-description} is then developed to enable the prover, possessing witness $(j,\mathbf{s},\mathbf{u}^{(1)}, \mathbf{u}^{(2)}, \mathbf{e}^{(1)}, \mathbf{e}^{(2)})$, to convince the verifier, with public input $(\Gmc^{(1)}, \Gmc^{(2)}, \mathbf{H}, \y_0, \ldots, \y_{N-1}, \mathbf{c}^{(1)}, \mathbf{c}^{(2)})$, that 
\begin{align}\label{eqn:CCA-relation-signing}
&\mathbf{H}\cdot \mathbf{s}^\top = \y_j^\top, \quad  
\mathbf{s} \in \mathsf{B}(m, \omega), \quad 
\mathbf{e}^{(1)} \in \mathsf{B}(\nmc, \tmc), \quad 
\mathbf{e}^{(2)} \in \mathsf{B}(\nmc, \tmc), \notag\\
&\big(\hspace*{1.5pt}\mathbf{u}^{(1)}\hspace*{1.5pt} \| \hspace*{1.5pt}\itob(j)\hspace*{1.5pt}\big)\cdot \Gmc^{(1)} \oplus \mathbf{e}^{(1)} = \mathbf{c}^{(1)}, \quad
\big(\hspace*{1.5pt}\mathbf{u}^{(2)}\hspace*{1.5pt} \| \hspace*{1.5pt}\itob(j)\hspace*{1.5pt}\big)\cdot \Gmc^{(2)} \oplus \mathbf{e}^{(2)} = \mathbf{c}^{(2)}.
\end{align}
The protocol employs the same technical ideas of the one described in Section~\ref{sec:nizk}. The facts that ciphertexts $\mathbf{c}^{(1)}$ and $\mathbf{c}^{(2)}$ encrypt the same plaintext $\itob(j)$ is proved in zero-knowledge by executing two instances of the techniques for handling one ciphertext $\mathbf{c}$ in the protocol of Section~\ref{sec:nizk}. The description of the protocol is given in Section~\ref{sec:appendix-CCA-ZK-protocol}.
	
Let $\Pi$ be the \textsf{NIZKAoK} obtained by repeating the protocol $\kappa=\omega \cdot \log{\lambda}$ times and making it non-interactive via the Fiat-Shamir heuristic. The group signature is set to be $\Sigma = (\mathbf{c}^{(1)}, \mathbf{c}^{(2)}, \Pi)$, where
\begin{equation}\label{eqn:CCA-NIZKAoK-Pi}
\hspace*{-15pt}\Pi = \left(\CMT^{(1)}, \ldots, \CMT^{(\kappa)}; \hspace*{2.5pt}(\Ch^{(1)}, \ldots, \Ch^{(\kappa)});\hspace*{2.5pt}\RSP^{(1)}, \ldots, \RSP^{(\kappa)}\right)
\end{equation}
and
$\left(\Ch^{(1)}\hspace*{-1pt}, \ldots, \Ch^{(\kappa)}\right) 
\hspace*{-1pt}=\hspace*{-1pt} \hash\left(M; \CMT^{(1)}\hspace*{-1pt}, \ldots, \CMT^{(\kappa)}; \mathsf{gpk}, \mathbf{c}^{(1)}, \mathbf{c}^{(2)}\right) \in \{1,2,3\}^\kappa$.

\smallskip
	
\item \textsf{Verify}$(\mathsf{gpk}, M, \Sigma)$: This algorithm proceeds as the verification algorithm of the scheme from Section~\ref{subsec:Scheme-description}, with $\left(\mathbf{c}^{(1)}, \mathbf{c}^{(2)}\right)$ taking the place of $\mathbf{c}$.
	
\smallskip
	
\item \textsf{Open}$(\mathsf{gmsk}, M, \Sigma)$: Parse $\Sigma$ as $\left(\mathbf{c}^{(1)}, \mathbf{c}^{(2)}, \Pi\right)$ and run $\mcdec\left(\mathsf{gmsk},\mathbf{c}^{(1)}\right)$ to decrypt $\mathbf{c}^{(1)}$. If decryption fails, then return $\bot$. If decryption outputs $\mathbf{g} \in \F_2^\ell$, then return $j = \btoi(\mathbf{g}) \in [0,N-1]$.
\end{enumerate}

The efficiency, correctness, and security aspects of the above group signature scheme are summarized in the following theorem.
\begin{theorem}
The given group signature scheme is correct. The public key has size $2nk + (m+N)r$ bits while the signatures have bit-size bounded above by $\big((N + 3\log N) + m(\log m +1) + 2n(\log n +1) + 2k + 5\lambda\big)\kappa +n$. In the random oracle model we can make two further assertions. First, if the Decisional McEliece problem $\mathsf{DMcE}(\nmc, \kmc, \tmc)$ and the Decisional Learning Parity with fixed-weight Noise problem $\mathsf{DLPN}(\kmc-\ell, \nmc, \mathsf{B}(\nmc, \tmc))$ are hard, and if the underlying $\mathsf{NIZKAoK}$ system is simulation-sound, then the scheme is $\mathsf{CCA}$-anonymous. Second, if the Syndrome Decoding problem $\mathsf{SD}(m, r, \omega)$ is hard, then the scheme is traceable.
\end{theorem}

Compared with the basic scheme in Section~\ref{sec:main-construction}, the present scheme introduces one more McEliece encrypting matrix of size $nk$ bits in the group public key, one more $n$-bit ciphertext and its supporting ZK sub-argument of well-formedness contained in $\Pi$ in the group signature. Overall, the upgrade from \textsf{CPA}-anonymity to \textsf{CCA}-anonymity incurs only a small and reasonable overhead in terms of efficiency. Since the correctness and traceability analyses of the scheme are almost identical to those of the basis scheme in Section~\ref{sec:main-construction}, the details are omitted here. In the next subsection, we will prove the \textsf{CCA}-anonymity property, which is the distinguished feature that we aim to accomplish.

\subsection{A Scheme Achieving CCA-Anonymity}\label{section:CCA}

Let $\mathcal{A}$ be any PPT adversary attacking the \textsf{CCA}-anonymity of the scheme with advantage $\epsilon$. We will prove that $\epsilon=\mathsf{negl}(\lambda)$ based on the \textsf{ZK} property and simulation-soundness of the underlying argument system. We keep the assumed hardness of the $\mathsf{DMcE}(\nmc, \kmc, \tmc)$ and the $\mathsf{DLPN}(\kmc-\ell, \nmc, \mathsf{B}(\nmc, \tmc))$ problems. Specifically, we consider the following sequence of hybrid experiments $G_0^{(b)}, G_1^{(b)}, \ldots, G_{10}^{(b)}, G_{11}$, where $b$ is the bit chosen by the challenger when generating the challenge signature. 

\smallskip

\noindent
\textbf{Experiment} $G_0^{(b)}$. This is the real \textsf{CCA}-anonymity experiment. The challenger runs $\mathsf{KeyGen}(1^\lambda, 1^N)$ to obtain 
\[
\big(\hspace*{2.5pt}\mathsf{gpk} = (\Gmc^{(1)}, \Gmc^{(2)}, \mathbf{H}, \y_0, \ldots, \y_{N-1}), \hspace*{5pt}\mathsf{gmsk} = \skmc^{(1)}, \hspace*{5pt}\mathsf{gsk}= (\mathsf{gsk}[0], \ldots, \mathsf{gsk}[N-1])\hspace*{2.5pt}\big),
\]
and then gives $\mathsf{gpk}$ and $\{\mathsf{gsk}[j]\}_{j \in [0, N-1]}$ to $\mathcal{A}$.  Queries to the opening oracle are answered using the opening secret key $\skmc^{(1)}$. 
In the challenge phase, $\mathcal{A}$ outputs a message $M^*$ together with two indices $j_0, j_1 \in [0, N-1]$. The challenger sends back a challenge signature $\Sigma^* = \left(\mathbf{c}^{(1), *}, \mathbf{c}^{(2), *}, \Pi^*\right) \leftarrow \mathsf{Sign}(\mathsf{gpk}, \mathsf{gsk}[j_b])$. The adversary outputs $b$ with probability $1/2 + \epsilon$.

\smallskip

\noindent
\textbf{Experiment} $G_1^{(b)}$. The only difference between this experiment and $G_0^{(b)}$ is that, when running $\mathsf{KeyGen}(1^\lambda, 1^N)$, the challenger retains the the second decryption key $\skmc^{(2)}$ instead of discarding it. The view of $\mathcal{A}$ in the two experiments are identical. 

\smallskip

\noindent
\textbf{Experiment} $G_2^{(b)}$. This experiment is like Experiment $G_1^{(b)}$ with one modification in the signature opening oracle. Instead of using $\skmc^{(1)}$ to open signatures, the challenger uses $\skmc^{(2)}$. It is easy to see that $\mathcal{A}$'s view will be the same as in Experiment $G_1^{(b)}$ until an event $F_2$ when $\mathcal{A}$ queries the opening of a signature $\Sigma = (\mathbf{c}^{(1)}, \mathbf{c}^{(2)}, \Pi)$ for which $\mathbf{c}^{(1)}$ and $\mathbf{c}^{(2)}$ encrypt distinct elements of $\mathbb{F}_2^\ell$. Since such an event $F_2$ could break the soundness of the zero-knowledge protocol used to generate $\Pi$, it could happen only with negligible probability. Therefore, the probability that $\mathcal{A}$ outputs $b$ in this experiment is negligibly close to $1/2 + \epsilon$. 

\smallskip

\noindent
\textbf{Experiment} $G_3^{(b)}$.  This experiment is identical to Experiment $G_2^{(b)}$, except on one modification. Instead of faithfully computing the \textsf{NIZKAoK} $\Pi^*$ using witness $(j,\mathbf{s},\mathbf{u}^{(1)}, \mathbf{u}^{(2)}, \mathbf{e}^{(1)}, \mathbf{e}^{(2)})$, the challenger simulates it by running the simulator of the underlying zero-knowledge protocol and programming the random oracle $\mathcal{H}$. Note that $\Pi^*$ is a simulated argument for a true statement, since $\mathbf{c}^{(1), *}$ and $\mathbf{c}^{(2), *}$ are honestly computed. Thanks to the statistical zero-knowledge property of the underlying protocol, Experiment $G_3^{(b)}$ is statistically close to Experiment $G_2^{(b)}$. 

\smallskip

\noindent
\textbf{Experiment} $G_4^{(b)}$. This experiment is similar to Experiment $G_2^{(b)}$ in the proof of CPA-anonymity in Section~\ref{subsec:proof-anonymity}. The only modification, with respect to the encrypting key $\Gmc^{(1)}$, is that instead of generating it using $\mckeygen(\nmc, \kmc, \tmc)$, we sample it uniformly at random over $\F_2^{\kmc\times\nmc}$. By the assumed hardness of the $\mathsf{DMcE}(\nmc, \kmc, \tmc)$ problem, this experiment is computationally indistinguishable from the previous experiment.

\smallskip

\noindent
\textbf{Experiment} $G_5^{(b)}$. This experiment is similar to Experiment $G_3^{(b)}$ in the proof of CPA-anonymity in Section~\ref{subsec:proof-anonymity}. Instead of computing $\mathbf{c}^{(1), *}$ as
\[
\mathbf{c}^{(1), *} = \big(\hspace*{1.5pt}\mathbf{u}^{(1)}\hspace*{1.5pt}\|\hspace*{1.5pt}\itob(j_b)\hspace*{1.5pt}\big)\cdot\Gmc^{(1)} \oplus \mathbf{e}^{(1)} = (\mathbf{u}^{(1)}\cdot \Gmc_1^{(1)} \oplus \mathbf{e}^{(1)}) \oplus \itob(j_b)\cdot\Gmc_2^{(1)},
\]
we let $\mathbf{c}^{(1),*} = \mathbf{v}^{(1)} \oplus \itob(j_b)\cdot\Gmc_2^{(1)}$, where $\mathbf{v}^{(1)} \xleftarrow{\$} \F_2^{\nmc}$. The simulated \textsf{NIZKAoK} $\Pi^*$ now corresponds to a false statement, since $\mathbf{c}^{(1),*}$ is not a well-formed ciphertext. Nevertheless, in the challenge phase, assuming the hardness of the $\mathsf{DLPN}(\kmc-\ell, \nmc, \mathsf{B}(\nmc, \tmc))$ problem, the adversary $\mathcal{A}$ can only observe the modification of $\mathbf{c}^{(1),*}$ with probability at most negligible in $\lambda$. The view of $\mathcal{A}$ is thus computationally close to that in Experiment $G_4^{(b)}$ above until an event $F_5$ which could happen after the challenge phase when $\mathcal{A}$ queries the opening of a signature $\Sigma = \left(\mathbf{c}^{(1)}, \mathbf{c}^{(2)}, \Pi\right)$ for which $\mathbf{c}^{(1)}$ and $\mathbf{c}^{(2)}$ encrypt distinct elements of $\mathbb{F}_2^\ell$. Since such an event $F_5$ could break the simulation-soundness of the underlying \textsf{NIZKAoK}, it could happen only with negligible probability. Therefore, the success probability of $\mathcal{A}$ in this experiment is negligibly close to that in Experiment $G_4^{(b)}$. 

\smallskip

\noindent
\textbf{Experiment} $G_6^{(b)}$. This experiment modifies Experiment $G_5^{(b)}$slightly. The ciphertext $\mathbf{c}^{(1),*}$ is now set as $\mathbf{c}^{(1),*} = \mathbf{r}^{(1)} \xleftarrow{\$} \F_2^{\nmc}$. Clearly, the distributions of $\mathbf{c}^{(1),*}$ in $G_5^{(b)}$ and $G_6^{(b)}$ are identical. Hence, the two experiments are statistically indistinguishable.

\smallskip

\noindent
\textbf{Experiment} $G_7^{(b)}$. This experiment switches the encrypting key $\Gmc^{(1)}$ back to an honest key generated by $\mckeygen(\nmc, \kmc, \tmc)$, and store the corresponding decryption key $\skmc^{(1)}$. By the assumed hardness of the $\mathsf{DMcE}(\nmc, \kmc, \tmc)$ problem, this experiment is computationally indistinguishable from $G_6^{(b)}$.

\smallskip

\noindent
\textbf{Experiment} $G_8^{(b)}$. In this experiment, we use $\skmc^{(1)}$, instead of $\skmc^{(2)}$, to answer signature opening queries. The view of $\mathcal{A}$ is identical to that in Experiment $G_7^{(b)}$ above until an event $F_8$ when $\mathcal{A}$ queries the opening of a signature $\Sigma = \left(\mathbf{c}^{(1)}, \mathbf{c}^{(2)}, \Pi\right)$ for which $\mathbf{c}^{(1)}$ and $\mathbf{c}^{(2)}$ encrypt distinct elements of $\mathbb{F}_2^\ell$. Since such an event $F_8$ could break the simulation-soundness of the underlying \textsf{NIZKAoK}, it could happen only with negligible probability.

\smallskip

\noindent
\textbf{Experiment} $G_9^{(b)}$. This experiment is similar to Experiment $G_4^{(b)}$ above. We merely replace the randomized McEliece encrypting key $\Gmc^{(2)}$ by a uniformly random matrix in $\F_2^{\kmc\times\nmc}$. By the assumed hardness of the $\mathsf{DMcE}(\nmc, \kmc, \tmc)$ problem, this experiment is computationally indistinguishable from Experiment $G_8^{(b)}$.

\smallskip

\noindent
\textbf{Experiment} $G_{10}^{(b)}$. This experiment is akin to Experiment $G_5^{(b)}$. The second ciphertext $\mathbf{c}^{(2), *}$ is now computed as $\mathbf{c}^{(2),*} = \mathbf{v}^{(2)} \oplus \itob(j_b)\cdot\Gmc_2^{(2)}$, where $\mathbf{v}^{(2)} \xleftarrow{\$} \F_2^{\nmc}$. Assuming the hardness of the $\mathsf{DLPN}(\kmc-\ell, \nmc, \mathsf{B}(\nmc, \tmc))$ problem, this experiment is computationally indistinguishable from $G_9^{(b)}$. 

\smallskip

\noindent
\textbf{Experiment} $G_{11}$. This experiment resembles Experiment $G_6^{(b)}$ above. The second ciphertext $\mathbf{c}^{(2),*}$ is now set as $\mathbf{c}^{(2),*} = \mathbf{r}^{(2)} \xleftarrow{\$} \F_2^{\nmc}$. Clearly, the distributions of $\mathbf{c}^{(2),*}$ in $G_{10}^{(b)}$ and $G_{11}$ are identical. As a result, the two experiments are statistically indistinguishable. Moreover, since $G_{11}$ no longer depends on the challenger's bit $b$,  the advantage of $\mathcal{A}$ in this experiment is $0$.

The above discussion shows that Experiments $G_0^{(b)}, \ldots, G_{10}^{(b)}, G_{11}$ are indistinguishable and that $\mathcal{A}$ has no advantage in game $G_{11}$. 
It then follows that the advantage of $\mathcal{A}$ in attacking the \textsf{CCA}-anonymity of the scheme, \ie, in Experiment $G_0^{(b)}$, is negligible. This concludes the justification for the \textsf{CCA}-anonymity property.

\section{Implementation Results}\label{sec:implement}
This section presents basic implementation results of our proposed group signature schemes to demonstrate their feasibility. 

\subsection{Test Environment}
The testing platform was a modern PC with a $3.4$~GHz Intel Core i5 CPU and $32$~GB of RAM. We employed the NTL~\cite{NTL} and $\mathsf{gf2x}$~\cite{GF2X} libraries for efficient polynomial operations over any field of characteristic $2$. The Paterson algorithm~\cite{Pat75} was used to decode binary Goppa codes in our implementation of the McEliece encryption. We employed SHA-3 with various output sizes to realize several hash functions. 

To achieve $80$-bit security, we chose the following parameters. The McEliece parameters were set to $(\nmc, \kmc, \tmc) = (2^{11}, 1696, 32)$, as in~\cite{BiswasS08}. The parameters for Syndrome Decoding were set to $(\ncfs, \kcfs, \tcfs) = (2756, 550, 121)$ so that the distribution of $\y_0, \ldots, \y_{N-1}$ is $2^{-80}$-close to the uniform distribution over $\F_2^r$, by Lemma~\ref{lemma:Leftover-Hash-Lemma}, and that the $\mathsf{SD}(m,r, \omega)$ problem is intractable with respect to the best known attacks. In particular, these parameters ensure the following work factor evaluations. First, the Information Set Decoding algorithm proposed in~\cite{BeckerJMM12} has work factor more than $2^{80}$. For an evaluation formula, one can also refer to~\cite[Slide~3]{Sendrier-workshop-slides}. Second, the birthday attacks presented in~\cite{FS09} have work factors more than $2^{80}$. The number of protocol repetitions $\kappa$ was set to $140$ to obtain soundness $1-2^{-80}$.

\subsection{Experimental Results}

Table~\ref{table:implementation} shows the implementation results of our \textsf{CPA}-anonymous group signature scheme, together with its public key and signature sizes, with respect to various numbers of group users and different message sizes. To reduce the signature size, in the underlying zero-knowledge protocol, we sent a random seed instead of permutations when $\text{Ch}=2$. Similarly, we sent a random seed instead of the whole response $\mathsf{RSP}$ when $\text{Ch}=3$. Using this technique, the average signature sizes were reduced to about $159$~KB for $4,096=2^{12}$ users and $876$~KB for $65,536 = 2^{16}$ users, respectively. Our public key and signature sizes are linear in the number of group users $N$, but it does not come to the fore while $N$ is less than $2^{12}$ due to the respective sizes of $\mathbf{G}$ and $\mathbf{H}$.

\begin{table*}[!ht]\centering
\renewcommand{\arraystretch}{1.2}
\caption{Implementation results and sizes of Our \textsf{CPA}-Anonymous Scheme. }\label{table:implementation}
\begin{tabular}{|c||c|c||c|c|c|c|c|}
\hline
\multirow{2}{*}{$N$} & \multirow{2}{*}{PK Size} & Average  & \multirow{2}{*}{Message} & \multirow{2}{*}{KeyGen} & \multirow{2}{*}{Sign} & \multirow{2}{*}{Verify} & \multirow{2}{*}{Open}\\
&		&	Signature Size & & & & & \\
\hline
\hline
$2^4$ & \multirow{2}{*}{625~KB} & \multirow{2}{*}{111~KB} & 1~~~B & \multirow{2}{*}{5.448} & 0.044 & 0.031 & \multirow{2}{*}{0.112} \\\cline{4-4}\cline{6-7}
(=16) & & & 1~GB & & 5.372 & 5.355 &  \\\hline
$2^8$ & \multirow{2}{*}{642~KB} & \multirow{2}{*}{114~KB} & 1~~~B & \multirow{2}{*}{5.407} & 0.045 & 0.032 & \multirow{2}{*}{0.111} \\\cline{4-4}\cline{6-7}
(=256) & & & 1~GB & & 5.363 & 5.351 &  \\\hline
$2^{12}$ & \multirow{2}{*}{906~KB} & \multirow{2}{*}{159~KB} & 1~~~B & \multirow{2}{*}{5.536} & 0.058 & 0.040 & \multirow{2}{*}{0.112} \\\cline{4-4}\cline{6-7}
(=4,096) & & & 1~GB & & 5.366 & 5.347 &  \\\hline
$2^{16}$ & \multirow{2}{*}{5.13~MB} & \multirow{2}{*}{876~KB} & 1~~~B & \multirow{2}{*}{7.278} & 0.282 & 0.186 & \multirow{2}{*}{0.111} \\\cline{4-4}\cline{6-7}
(=65,536) & & & 1~GB & & 5.591 & 5.497 &  \\\hline
$2^{20}$ & \multirow{2}{*}{72.8~MB} & \multirow{2}{*}{12.4~MB} & 1~~~B & \multirow{2}{*}{33.947} & 3.874 & 2.498 & \multirow{2}{*}{0.111} \\\cline{4-4}\cline{6-7}
(=1,048,576) & & & 1~GB & & 9.173 & 7.795 &  \\\hline
$2^{24}$ & \multirow{2}{*}{1.16~GB} & \multirow{2}{*}{196~MB} & 1~~~B & \multirow{2}{*}{481.079} & 61.164 & 39.218 & \multirow{2}{*}{0.111} \\\cline{4-4}\cline{6-7}
(=16,777,216) & & & 1~GB & & 66.613 & 44.575 &  \\\hline
\multicolumn{8}{l}{The unit for time is second. All implementation results are the averages from 100 tests.}
\end{tabular}
\end{table*}

Our implementation took $0.282$ and $0.186$ seconds for a $1$~B message and $5.591$ and $5.497$ seconds for a $1$~GB message, respectively, to sign a message and to verify a generated signature for a group of $65,536$ users. In our experiments, it takes about $5.30$ seconds to hash a $1$~GB message and it leads to the differences of signing and verifying times between a $1$~B and a $1$~GB messages.
One may naturally expect that running times should be increased once $N$ becomes larger. 
But, in Table II, the increases are negligible and on occasions the running time even decreases slightly as $N$ grew up to $2^{12}$. This could also be due to the effect that the time required to perform other basic operations with parameters $\mathbf{G}$ and $\mathbf{H}$ had on the overall running time.

\begin{table*}
\centering
\renewcommand{\arraystretch}{1.2}
\caption{Implementation results and sizes of Our \textsf{CCA}-Anonymous Scheme.}\label{table:implementation-cca}
\begin{tabular}{|c||c|c||c|c|c|c|c|}
\hline
\multirow{2}{*}{$N$} & \multirow{2}{*}{PK Size} & Average  & \multirow{2}{*}{Message} & \multirow{2}{*}{KeyGen} & \multirow{2}{*}{Sign} & \multirow{2}{*}{Verify} & \multirow{2}{*}{Open}\\
&		&	Signature Size & & & & & \\
\hline
\hline
$2^4$ & \multirow{2}{*}{1.06~MB} & \multirow{2}{*}{157~KB} & 1~~~B & \multirow{2}{*}{10.660} & 0.065 & 0.046 & \multirow{2}{*}{0.111} \\\cline{4-4}\cline{6-7}
(=16) & & & 1~GB & & 5.366 & 5.351 &  \\\hline
$2^8$ & \multirow{2}{*}{1.08~MB} & \multirow{2}{*}{160~KB} & 1~~~B & \multirow{2}{*}{10.605} & 0.066 & 0.046 & \multirow{2}{*}{0.112} \\\cline{4-4}\cline{6-7}
(=256) & & & 1~GB & & 5.382 & 5.369 &  \\\hline
$2^{12}$ & \multirow{2}{*}{1.34~MB} & \multirow{2}{*}{205~KB} & 1~~~B & \multirow{2}{*}{10.731} & 0.080 & 0.056 & \multirow{2}{*}{0.112} \\\cline{4-4}\cline{6-7}
(=4,096) & & & 1~GB & & 5.381 & 5.362 &  \\\hline
$2^{16}$ & \multirow{2}{*}{5.56~MB} & \multirow{2}{*}{922~KB} & 1~~~B & \multirow{2}{*}{12.438} & 0.309 & 0.202 & \multirow{2}{*}{0.111} \\\cline{4-4}\cline{6-7}
(=65,536) & & & 1~GB & & 5.629 & 5.519 &  \\\hline
$2^{20}$ & \multirow{2}{*}{73.2~MB} & \multirow{2}{*}{12.5~MB} & 1~~~B & \multirow{2}{*}{39.099} & 3.998 & 2.504 & \multirow{2}{*}{0.111} \\\cline{4-4}\cline{6-7}
(=1,048,576) & & & 1~GB & & 9.322 & 7.829 &  \\\hline
$2^{24}$ & \multirow{2}{*}{1.16~GB} & \multirow{2}{*}{196~MB} & 1~~~B & \multirow{2}{*}{490.219} & 62.878 & 39.358 & \multirow{2}{*}{0.111} \\\cline{4-4}\cline{6-7}
(=16,777,216) & & & 1~GB & & 68.177 & 44.648 &  \\\hline
\multicolumn{8}{l}{The unit for time is second. All implementation results are the averages from 100 tests.}
\end{tabular}
\end{table*}

Table~\ref{table:implementation-cca} contains the implementation results of our \textsf{CCA}-anonymous group signature scheme, along with public key and average signature sizes for various number of group users and different message sizes. 
The public key size of our \textsf{CCA}-anonymous scheme is $434$~KB larger than that of our \textsf{CPA}-anonymous version because it additionally requires the matrix $\mathbf{G}^{(2)}$. The average signature size is also about 46 KB larger since the response $\mathsf{RSP}$ additionally includes $\mathbf{v}_{\mathbf{e}}^{(2)}$, $\mathbf{w}_{\mathbf{e}}^{(2)}$ when $\text{Ch}=1$ and $\mathbf{z}_{\mathbf{u}}^{(2)}$, $\mathbf{z}_\mathbf{e}^{(2)}$ when $\text{Ch}=2$. We remark that there is no additional element to be sent for $\text{Ch}=3$ since we just sent a random seed instead of the whole response, as in the implementation of our \textsf{CPA}-anonymous scheme.
The results in Table~\ref{table:implementation-cca} show that the \textsf{CCA}-anonymous version requires only a small overhead for key generation, signing, and verification. For example, the key generation algorithm took about $5$ seconds more, which corresponds to the key generation time for the McEliece encryption. When the number of group users is less than $2^{20}$, it also took about $0.020$~seconds and $0.015$~seconds more to generate a signature and verify it, respectively. The overheads for signing are increased slightly once the number of group users is $2^{24}$, but they account for about $2.28\%$ and $2.73\%$ of total signing times for $1$~B and $1$~GB messages, respectively. 

In conclusion, to our best knowledge, the implementation results presented here are the first ones for {group signatures from quantum-resistant assumptions}. 
We have thus demonstrated that our schemes, while not yet truly practical, are bringing this class of group signatures closer to practice.

\section{Conclusion}\label{sec:conclude}

We put forward two provably secure code-based group signature schemes in the random oracle model. The first scheme satisfies the \textsf{CPA}-anonymity and traceability requirements for group signatures under the assumed hardness of three well-known problems in code-based cryptography. These are the McEliece problem, the Learning Parity with Noise problem and a variant of the Syndrome Decoding problem. We extend the basic scheme to achieve \textsf{CCA}-anonymity by exploiting the Naor-Yung transformation. 
The feasibility of the proposed schemes is backed by implementation results. 

The work we presented here inaugurates a foundational step in code-based group signatures. The natural continuation is to work towards either one of the following goals: {constructing practically efficient schemes whose signature sizes are sub-linear in the number of group users and obtaining provably secure schemes in the QROM or in the standard model}.

\section*{Acknowledgements}

The authors would like to thank Jean{-}Pierre Tillich, Philippe Gaborit, Ayoub Otmani, 
Nicolas Sendrier, and Nico D{\"{o}}ttling for helpful comments and discussions. 

\end{document}